\documentclass[11pt,a4paper]{article}
\pdfoutput=1

\usepackage{bm}
\usepackage{amsmath,amssymb}
\usepackage[makeroom]{cancel}
\usepackage{cite}

\usepackage{graphicx}
\usepackage{color}
\usepackage{colortbl}
\usepackage{multirow}

\usepackage{tikz}
\usetikzlibrary{patterns}

\usepackage{hyperref} 
\hypersetup{colorlinks=true, citecolor=blue, filecolor=black, linkcolor=blue, urlcolor=blue, pdfpagemode=UseNone}

\numberwithin{figure}{section}
\numberwithin{equation}{section}

\newcommand\ie{\textit{i.e.}\ }
\newcommand\eg{\textit{e.g.}\ }
\newcommand\cf{\textit{cf.}\ }

\newcommand{\be}{\begin{equation}}
\newcommand{\ee}{\end{equation}}
\newcommand{\bea}{\begin{eqnarray}}
\newcommand{\eea}{\end{eqnarray}}

\newcommand{\eps}{\varepsilon}
\newcommand{\vp}{\varphi}  
\newcommand{\bp}{\bar \varphi} 
\newcommand{\bc}{\bar \chi} 
\newcommand{\bV}{\bar V} 
\newcommand{\bg}{\bar \gamma} 

\newcommand{\hV}{\hat V} 
\newcommand{\hp}{\hat \phi} 
\newcommand{\hk}{\hat k} 
\newcommand{\hc}{\hat \chi} 
\newcommand{\hatt}{\hat t} 

\newcommand{\dclnf}{\,\partial_\chi\! \ln\! f \,}

\begin{document}

\vspace*{2cm}
\begin{center}
{\huge \bf Background independence in a background dependent renormalization group}
\vspace*{0.1cm}
\vskip 1cm

\renewcommand{\thefootnote}{\fnsymbol{footnote}}
\setcounter{footnote}{1}

{\large  Peter Labus, $^a$\footnote{plabus@sissa.it} Tim\ R.\ Morris $^b$\footnote{T.R.Morris@soton.ac.uk} and Zo\"e H. Slade $^b$\footnote{Z.Slade@soton.ac.uk}} 
\vskip 8pt

{$^a$ \it International School for Advanced Studies, via Bonomea 265, 34136 Trieste, Italy \\
              and INFN, Sezione di Trieste}\\
{$^b$  \it STAG Research Centre \& School of Physics and Astronomy,\\
University of Southampton,
              Highfield, Southampton, SO17 1BJ, U.K.}\\

\vspace*{0.5cm}

\end{center}


\begin{center}
{\large \bf Abstract}

\vspace{.6cm}

\parbox{14cm}{
Within the derivative expansion of conformally reduced gravity, the modified split Ward identities are shown to be compatible with the flow equations if and only if either the anomalous dimension vanishes or the cutoff profile is chosen to be power law. No solutions exist if the Ward identities are incompatible. In the compatible case, a clear reason is found for why Ward identities can still forbid the existence of fixed points; however, for any cutoff profile, a background independent (and parametrisation independent) flow equation is uncovered. Finally, expanding in vertices, the combined equations are shown generically to become either over-constrained or highly redundant beyond the six-point level.
}

\end{center}

\vspace{2cm}

\renewcommand{\thefootnote}{\arabic{footnote}}
\setcounter{footnote}{0}

\newpage

\tableofcontents

\newpage

\section{Introduction}\label{sec:introduction}

Following ref. \cite{Reuter:1996}, non-perturbative RG (renormalization group) flows for quantum gravity have been formulated by utilising the technical device of splitting the metric in terms of a background metric and a fluctuation field (see also the reviews \cite{Reuter:2012,Percacci:2011fr,Niedermaier:2006wt,Nagy:2012ef,Litim:2011cp}). Physical results must then be independent of this split, in other words should be background independent. The background metric however is in particular also used (via the background Laplacian) to define the effective cutoff scale $k$. This breaks background independence at intermediate scales $k$ but such that background independence can be recovered in the limit $k\to0$ providing certain modified split Ward identities (msWIs) are imposed \cite{Pawlowski:2005xe,Litim:2002hj,Bridle:2013sra,Reuter:1997gx,Litim:1998nf,Litim:2002ce,Manrique:2009uh,Manrique:2010mq,Manrique:2010am,Dietz:2015owa,Safari:2015dva}.

An unsettling conclusion from the research reported in ref. \cite{Dietz:2015owa} is that the requirement of background independence in a theory of quantum gravity can actually be in conflict with renormalization group (RG) properties in flows formulated as above: fixed points under changes in the effective cutoff scale $k$, can be forbidden by the msWIs  that are enforcing background independence.\footnote{Very recently an alternative approach has been initiated which avoids these issues entirely since background independence is never broken \cite{Morris:2016nda}.} 

In the conformally truncated gravity model investigated in ref.  \cite{Dietz:2015owa}, this happens generically when the anomalous dimension $\eta$ is non-vanishing.
It can however be avoided by a careful choice of parametrisation $f$ (setting it to be a power of $\chi$ determined by its scaling dimension \cite{Dietz:2015owa}).
On the other hand it was shown in ref. \cite{Dietz:2015owa} that the situation is saved in all cases, at least in the conformally reduced gravity model, by the existence of an alternative background-independent description. This involves in particular a background-independent notion of scale, $\hat{k}$. This background independent description exists at a deeper underlying level since in terms of these background-independent variables, the RG fixed points and corresponding flows always exist, and are manifestly independent of the choice of parametrisation $f(\chi)$.  

After approximating the exact RG flow equations and msWIs to second order in the derivative expansion (as will be reviewed later), 
the crucial technical insight was to notice that, just as in the scalar field theory model \cite{Bridle:2013sra}, the msWIs and RG flow equations can be combined into linear partial differential equations. It is the solution of the latter equations by the method of characteristics, that uncovers the background independent variables. And it is by comparing the description in these variables with the equivalent description in the original variables,
that we see that fixed points in the original variables are in general forbidden by  background independence.

However in order to facilitate combining the RG flow equations and msWIs when the anomalous dimension $\eta\ne0$, the authors of ref. \cite{Dietz:2015owa} were led to a particular form of cutoff profile $R_k$, namely a power-law cutoff profile. We will show in this paper that in fact this cutoff profile plays a r\^ole that is much deeper than the convenience of this mathematical trick. 
Whilst at the exact level the msWIs are guaranteed to be compatible with the exact RG flow equation (for completeness and later purposes we provide a proof in sec. \ref{sec:exact}), this will typically not be the case once approximated. 
We will see that in the $O(\partial^2)$ derivative expansion approximation derived in ref. \cite{Dietz:2015owa}, the msWI and flow equations are in fact compatible {if and only if} either the cutoff profile is power law, or we have the special case that $\eta=0$.

If the msWIs are not compatible with the flow equations, it does not immediately follow that there are no simultaneous solution to the system of equations. However, as we argue in sec. \ref{sec:incompatibility} and verify by example in sec. \ref{sec:incompatible-no-solns} (see also sec. \ref{sec:truncations}), if the msWIs are not compatible, the equations are overconstrained and it is for this reason that it is hopeless to expect any solutions.

The example is furnished by choosing optimised cutoff profile and LPA, but keeping $\eta$ non-vanishing.
In fact it is natural to expect $\eta$ to be non-vanishing at the LPA level for conformally truncated gravity, as explained in ref. \cite{Dietz2016}.  Choosing a power-law form for $f$ the equations at first sight then look consistent and able to support fixed points. Although we already know that the msWI is in this case incompatible with the flow equation, it is still possible to combine the msWI and flow equation into a linear partial differential equation. Solving this, we confirm that for the combined system there are no solutions with $\eta\ne0$, supporting the arguments in sec. \ref{sec:incompatibility}. In sec. \ref{sec:truncations}, by considering the simplest polynomial truncation, we also verify very straightforwardly that there can be no fixed points. 

For power-law cutoff profile the equations are compatible, however we will see very clearly in sec. \ref{sec:forbids} why with $\eta\ne0$ and non-power-law parametrisation $f$, there can be no fixed points with respect to $k$.

Actually, power law cutoff profiles have nice properties in  that they ensure that the derivative expansion approximation preserves the quantisation of the anomalous dimension in non-gravitational systems, \eg scalar field theory \cite{Morris:1994ie,Morris:1994jc,Morris:1998}.\footnote{Although as with the optimised cutoff \cite{opt1,opt3}, they do not allow a derivative expansion to all orders \cite{Morris:2005ck,Morris1999,Morris2001}.} Nevertheless, given the unsettling nature of the conclusions in ref. \cite{Dietz:2015owa}, it is important to understand  to what extent the results depend on cutoff profile. 

Since $\eta=0$ is sufficient to allow the derivative expansions of the exact RG and msWI equations to be compatible, we can therefore investigate in this case the implications of background independence, whatever the cutoff profile. The conclusions are the same for any cutoff profile that is not power law, so for simplicity and to make the equations completely explicit, we take the popular and simple choice of optimised cutoff \cite{opt1,opt3}.  Just as found for power law cutoff \cite{Dietz2016} in this case, background-independent variables exist, and $k$-fixed points exist; these coincide with the fixed points in background-independent variables.  In terms of background-independent variables and at LPA level, the full line of fixed points
is visible, confirming the findings for power-law cutoff \cite{Dietz2016}.

Although we can show this by exact analysis of the LPA flow and msWI equations, using the trick of combining these equations into a linear partial differential equation, it is very instructive to analyse these equations without using this trick and also by considering only polynomial truncations, since it seems likely that this is the only way we could investigate this issue using the exact non-perturbative flow equations. Viewed from this perspective, we will see that the problem is that if the RG fixed point equations and msWI equations are truly independent, then they will overconstrain the solutions if carried to sufficiently high order truncation. Indeed, expanding in powers of the fluctuation field $\vp$ to the $m^{\rm th}$ level and background field $\chi$ to the $n^{\rm th}$ level, we get one fixed point equation for each $(m,n)$-point vertex \emph{and} one msWI equation per vertex. Even though each of these equations is open (depending on yet higher-point vertices) we will see that since there are two equations for every vertex, at sufficiently high order truncation there are more equations than vertices (indeed eventually double the number) and thus either the equations become highly redundant or the vertices are constrained to the point where there are no solutions.

This analysis strongly suggests therefore that the full non-perturbative Ward identities would lead to important constraints on RG properties. Unfortunately it seems very challenging to investigate this, since we will see that the number of equations only exceeds the number of vertices for the first time  at the six-point level.
We discuss potential conflict for the exact non-perturbative flow equations further in the conclusions.

\section{Conformally reduced gravity at order derivative-squared}\label{sec:review}

In this section we give a quick resum\'e of the results we need and their context, from ref. \cite{Dietz:2015owa}.
%
We arrive at conformally reduced gravity (in Euclidean signature) by writing:
\begin{equation}
\label{conformal-reduction}
	\tilde g_{\mu\nu} = f(\tilde\phi) \hat g_{\mu\nu}= f(\chi +\tilde\varphi )\hat g_{\mu\nu} \qquad\text{and} \qquad \bar g_{\mu\nu} = f(\chi)\hat g_{\mu\nu} \,.
\end{equation}
Here $\tilde g_{\mu\nu}$ is the metric that is integrated over in the partition function. It is restricted to an overall conformal factor $f(\tilde\phi)$ times a fiducial metric which in fact we set to flat: $\hat g_{\mu\nu}=\delta_{\mu\nu}$. 

Examples of parametrisations used previously in the literature include  
$f(\phi) = \exp(2\phi)$ \cite{Machado:2009ph} and $f(\phi)=\phi^2$ \cite{Manrique:2009uh,Bonanno:2012dg}. However we leave the choice of parametrisation $f$ unspecified. It is important to note however that $f$ cannot depend on $k$ since it is introduced at the bare level and has no relation to the infrared cutoff (moreover if $f$ depended on $k$, the flow equation \eqref{equ:flowGamma} would no longer hold).  Later we will change to dimensionless variables using $k$ and in these variables it can be forced to depend on $k$   (see especially secs. \ref{sec:required-cutoff} and \ref{sec:forbids}).

We split the total conformal factor field $\tilde\phi(x)$ into a background conformal factor field $\chi(x)$ and fluctuation conformal factor field $\tilde\vp(x)$. It is then the latter that is integrated over. Also as shown, we similarly parametrise the background metric $\bar{g}_{\mu\nu}$ in terms of the background conformal factor field $\chi$. Introducing the classical fluctuation field $\vp = \langle \tilde \vp \rangle$ and total classical field $\phi = \langle \tilde \phi \rangle = \chi + \vp$, the effective action  satisfies the flow equation
\be
\label{equ:flowGamma}
\partial_t \Gamma_k[\vp,\chi] = \frac{1}{2}\mathrm{Tr}\left[\frac{1}{\sqrt{\bar g}\sqrt{\bar g}}\frac{\delta^2\Gamma_k}
				  {\delta \vp \delta \vp}+ R_k[\chi]\right]^{-1} \partial_t R_k[\chi]\,.
\ee
 Here we have introduced the RG time 
\be
\label{time}
t=\ln(k/\mu)\,,
\ee 
with $\mu$ being a fixed reference scale, which can be thought of as being the usual arbitrary finite physical mass-scale. $R_k$ is the cutoff operator responsible for suppressing momentum modes below the infrared cutoff scale $k$, \cf \cite{Wetterich:1992,Morris:1993}. 
%
The crucial observation is that in the context of the background field method in quantum gravity the cutoff operator itself depends on the background field $\chi$. The reason for this is that the cutoff operator is a function of the covariant Laplacian of the background metric $R_k\left(-\bar \nabla^2\right)$, as it is with respect to the spectrum of $-\bar\nabla^2$ that modes are integrated out or suppressed in the path integral, \cf \cite{Reuter:2008wj,Reuter:2009kq}. 

A remnant diffeomorphism invariance enforces this $\chi$ dependence in the approximation chosen in ref. \cite{Dietz:2015owa}.
By specialising to a background metric ${\bar g}_{\mu\nu}$ that is slowly varying, so that space-time derivatives of this can be neglected, we
effectively terminate at the level of the LPA for the background conformal factor $\chi$. For the classical fluctuating conformal factor $\vp$ however,  $\mathcal{O}(\partial^2)$ in the derivative expansion approximation is fully implemented, making no other approximation.
The effective action thus takes its most general form at this level of truncation:
\begin{equation}
\label{trunc}
	\Gamma_k[\varphi, \chi] = \int d^dx \sqrt{\bar g} \left( -\frac{1}{2}K(\varphi,\chi)
	\bar g^{\mu\nu}\partial_{\mu}\varphi\partial_{\nu}\varphi + V(\varphi,\chi)  \right)\,.
\end{equation}
The msWI encodes the extent to which the effective action violates split symmetry:
\be
\label{equ:split-symmetry}
\tilde \vp(x) \mapsto \tilde \vp(x) + \eps(x) \qquad \chi(x) \mapsto \chi(x) -\eps(x)\,.
\ee
Due to the special r\^ole played by $\chi$, the infrared cutoff operator breaks this symmetry, leading to the msWI:
\be
\label{equ:sWiGamma}
\frac{1}{\sqrt{\bar g}}\left(\frac{\delta\Gamma_k}{\delta \chi}-\frac{\delta \Gamma_k}{\delta \vp}\right)
      =\frac{1}{2}\mathrm{Tr}\left[\frac{1}{\sqrt{\bar g}\sqrt{\bar g}}\frac{\delta^2\Gamma_k}
				  {\delta \vp \delta \vp}+ R_k[\chi]\right]^{-1} \frac{1}{\sqrt{\bar g}}
				  \left\{\frac{\delta R_k[\chi] }{\delta \chi}+\frac{d}{2}\dclnf R_k[\chi]\right\}\,.
\ee
Exact background independence would be realised if the right hand side of the msWI was zero, implying that the effective action is only a functional of the total field $\phi = \chi + \vp$. The presence of the cutoff operator however causes the right hand side to be non-vanishing in general. It is only in the limit $k\rightarrow0$ (holding physical, \ie unscaled, momenta and fields fixed) that the cutoff operator drops out and background independence can be restored exactly. We note therefore that imposing the msWI in addition to the flow equation \eqref{equ:flowGamma} automatically ensures exact background independence in the limit $k\rightarrow0$. The observation we further explore in this paper is that restricting flows to satisfy \eqref{equ:sWiGamma} then has consequences for the RG properties, in particular fixed point behaviour, that follows from \eqref{equ:flowGamma}.


Computing the flow equation and msWI in the derivative expansion \eqref{trunc} results in flow equations and modified split Ward identities\footnote{Although we always mean these modified identities, we will sometimes refer to them  simply as Ward identities.},  for the potential $V$:
\begin{align}
	\label{flowV}
	\partial_t V(\varphi,\chi) &= f(\chi)^{-\frac{d}{2}}\int dp\, p^{d-1} Q_p\dot R_p \,,\\
	\label{msWIV}
	\partial_\chi V - \partial_\varphi V +\frac{d}{2}\partial_\chi \text{ln} f V 
	&= f(\chi)^{-\frac{d}{2}}\int dp\, p^{d-1} Q_p\left(\partial_\chi R_p + \frac{d}{2}\partial_\chi \text{ln} f  R_p\right),
\end{align}
and for $K$:
\begin{align}
	\label{flowK}
	f^{-1}\partial_t K(\varphi,\chi) &= 2 f^{-\frac{d}{2}}\int dp\,p^{d-1} P_p(\varphi,\chi)\dot R_p \,,\\
	\label{msWIK}
	 f^{-1}\left(\partial_\chi K- \partial_\varphi K + \frac{d-2}{2}\partial_\chi\text{ln}fK\right)
	 &= 2 f^{-\frac{d}{2}}\int dp\,p^{d-1} P_p(\varphi,\chi) \left(\partial_{\chi}R_p 
	 + \frac{d}{2}\partial_\chi \text{ln} f   R_p\right).
\end{align}
The $p$ subscripts denote the momentum dependence of $Q_p, P_p$ and the cutoff $R_p$ and as usual RG time derivatives are denoted also by a dot on top. $Q_p$ is defined as
\begin{equation}
	\label{Q}
	Q_p=\left(\partial^2_\varphi V - p^2\frac{K}{f} + R_p \right)^{-1}.
\end{equation}
and $P_p$ is given by
\begin{align}
	P_p&=-\frac{1}{2}\frac{\partial_\varphi K}{f}Q_p^2
	+\frac{\partial_\varphi K}{f}\left(2\partial_\varphi^3 V 
	- \frac{2d+1}{d}\frac{\partial_{\varphi}K}{f}p^2\right)Q_p^3\nonumber\\
	&-\left[\left\{\frac{4+d}{d}\frac{\partial_\varphi K}{f}p^2 - \partial^3_\varphi V\right\}
	\left(\partial_{p^2}R_p - \frac{K}{f}\right) 
	+\frac{2}{d}p^2\partial^2_{p^2} R_p\left(\frac{\partial_\varphi K}{f} - \partial_\varphi^3V\right)\right]
	\left(\partial_\varphi^3 V - \frac{\partial_\varphi K}{f}p^2\right)Q_p^4\nonumber\\
	&-\frac{4}{d}p^2\left(\partial_{p^2}R_p-\frac{K}{f}\right)^2
	\left(\partial_\varphi^3 V - \frac{\partial_\varphi K}{f} p^2\right)^2 Q_p^5 \,.
\end{align}

\section{Compatibility of the msWI with the flow equation}

Compatibility of the msWI with the flow equation means the following. Write the msWI in the form $\mathcal{W}=0$ and assume that this holds at some scale $k$. Computing $\dot{\mathcal{W}}$ by using the flow equation, we say that the msWI is compatible if $\dot{\mathcal{W}}=0$ then follows at scale $k$ without further constraints. 

In the first part of this section we rederive the flow equation and msWI for conformally reduced gravity but organised in a different way from ref. \cite{Dietz:2015owa} so as to make the next derivation more transparent. We then prove that they are compatible with one another. So far, this is naturally to be expected since both are derived from the same partition function. For completeness we include it here in order to  fully understand the issues once we consider derivative expansions. (For a proof of the exact case in a more general context see ref. \cite{Safari:2015dva}.)
In the second part we study the notion of compatibility for conformally reduced gravity in the truncation \eqref{trunc}. Asking for compatibility in the derivative expansion is actually non-trivial. We derive the requirements necessary to achieve it.

\subsection{Compatibility at the exact level}\label{sec:exact}

The proof of compatibility of the un-truncated system consists of demonstrating that the RG time derivative of the msWI is proportional to the msWI itself \cite{Litim1998,Litim1999}. In analogy with references \cite{Litim1998,Litim1999}, we expect to find that this RG time derivative is, more specifically, proportional to a second functional derivative with respect to $\varphi$ acting on the msWI and it is with this in mind that we proceed (see also ref.\cite{Safari:2015dva}).

We begin by considering the following Euclidean functional integral over the fluctuation field $\tilde\varphi$
\begin{equation}
	\label{Z}
	\text{exp}(W_k)=\int\!\mathcal{D}\tilde\varphi \, \text{exp}\left(-S[\chi+\tilde\varphi]-S_k[\tilde\varphi,\bar g]
	+ S_{\text{src}}[\tilde\varphi,\bar g]\right).
\end{equation}
This integral is regulated in the UV (as it must be), however we leave this regularisation implicit in what follows. Compatibility can be shown most easily by presenting both the flow equation and the msWI as matrix expressions. Thus we begin by rewriting the source term using matrix notation like so
\begin{equation}
	S_{\rm src}[\tilde\varphi,\bar g]=\int\!d^d x\sqrt{\bar g(x)}\,\tilde\varphi(x) J(x)\equiv 
	\tilde\varphi_xT_{xy}J_y\equiv \tilde\varphi\cdot T\cdot J \,,
\end{equation}
where $T_{xy}\equiv T(x,y)\equiv\sqrt{\bar g(x)}\delta(x-y)$ and the dot notation represents integration over position space.  Similarly, we write the cutoff action as
\begin{equation}
\label{cutoff-action}
	S_k[\tilde\varphi, \bar g]=\frac{1}{2}\int\!d^d x\sqrt{\bar g(x)}\,\tilde\varphi(x)R_k[\bar g]\tilde\varphi(x)
	\equiv\frac{1}{2}\tilde\varphi_x r_{xy} \tilde\varphi_{y}\equiv\frac{1}{2}\,\tilde\varphi\cdot r \cdot\tilde\varphi \,,
\end{equation}
where 
\be 
\label{odd-r}
r_{xy}\equiv r(x,y)\equiv\sqrt{\bar g(x)}\sqrt{\bar g(y)}R_{k}(x,y)\,,
\ee 
and where the cutoff operator and its kernel are related according to
\begin{equation}
	R_k(x,y) = R_{k,x}\frac{\delta(x-y)}{\sqrt{\bar g(y)}} \,.
\end{equation}
We refrain from putting a $k$ subscript on $r_{xy}$ to avoid clutter with indices, but note that it still has $k$-dependence. Also note that now the factors of $\sqrt{\bar g}$ are no longer part of the integration; this is to enable all $\chi$-dependent quantities to be easily accounted for when acting with $\delta/\delta\chi$ later on. With these definitions in place, the RG time derivative of \eqref{Z} gives
\begin{equation}
	\label{W_flow}
	\dot W_k=-\frac{1}{2}\dot r_{xy} \left<\tilde\varphi_x \tilde\varphi_y \right>.
\end{equation}
In the usual way, we take the Legendre transform of $W_k$:
\begin{equation}
	\label{LTR}
	\tilde\Gamma_{k}=J\cdot T \cdot \varphi - W_k \quad \text{with} \quad T\cdot\varphi=\frac{\delta W_k}{\delta J}
\end{equation}
and from this we define the effective average action
\begin{equation}
	\label{EAA}
 	\Gamma_k[\varphi,\bar g]=\tilde\Gamma_k[\varphi,\bar g]-S_k[\varphi,\bar g] \,.
\end{equation}
From \eqref{LTR}, it also follows that
\begin{equation}
	\label{2pt}
	\langle\tilde\varphi_x \tilde\varphi_y \rangle =
	\left(\frac{\delta^{2}\tilde\Gamma_k}{\delta\varphi_x \delta\varphi_y}\right)^{\!-1}+\,\varphi_x \varphi_y \,.
\end{equation}
Finally substituting \eqref{LTR} and \eqref{2pt} into \eqref{W_flow}, together with \eqref{EAA}, we obtain the flow equation for the effective average action
\begin{equation}
	\label{flow1}
	\dot\Gamma_k=\frac{1}{2}\text{tr}\left[\left(\frac{\delta^{2}\Gamma_k}{\delta\varphi \delta\varphi}
	+r\right)^{\!-1}\dot r\right]
	\equiv \frac{1}{2}\text{tr}\,\Delta\, \dot r \,,
\end{equation}
where
\begin{equation}
\label{new-inverse}
	\Delta_{xy}\equiv\left(\frac{\delta^{2}\Gamma_k}{\delta\varphi_x \delta\varphi_y}+r_{xy}\right)^{-1} \,.
\end{equation}
The msWI is derived by applying the  split symmetry transformations \eqref{equ:split-symmetry}, with infinitesimal $\varepsilon(x)$, to the functional integral \eqref{Z}.
The bare action is invariant under this shift, however the source term and cutoff action are not. It is the breaking of this symmetry that indicates background independence has been lost. Applying these shifts to \eqref{Z} we obtain
\begin{equation}
	\label{Wshift}
	-\frac{\delta W_k}{\delta\chi}\cdot\varepsilon=\left< \varepsilon\cdot T\cdot J-
	\tilde\varphi\cdot\left(\frac{\delta T}{\delta\chi}\cdot\varepsilon\right)\cdot J 
	- \varepsilon\cdot r\cdot\tilde\varphi
	+\frac{1}{2}\tilde\varphi\cdot\left(\frac{\delta r}{\delta\chi}\cdot\varepsilon\right)\cdot\tilde\varphi\right>.
\end{equation}
Under these same shifts, the Legendre transformation \eqref{LTR} gives
\begin{equation}
	\frac{\delta W_k}{\delta\chi}\cdot\varepsilon
	= J\cdot\left(\frac{\delta T}{\delta\chi}\cdot\varepsilon\right)\cdot\varphi
	- \frac{\delta\tilde\Gamma_k}{\delta\chi}\cdot\varepsilon \,.
\end{equation} 
Substituting the above relation into \eqref{Wshift} together with \eqref{EAA}, we obtain the msWI:
\begin{equation}
	\label{msWIuntrunc}
	\frac{\delta \Gamma_k}{\delta\chi_\omega}-\frac{\delta \Gamma_k}{\delta\varphi_\omega}=
	\frac{1}{2}\,\Delta_{xy}\,\frac{\delta r_{yx}}{\delta\chi_\omega} \,,
\end{equation}
where we have used the fact that the identity must hold for arbitrary $\varepsilon(\omega)$. Note that in deriving \eqref{msWIuntrunc} the contribution of the source term to the separate background field dependence of $\Gamma_k[\varphi,\chi]$ drops out.

The equations just derived, \eqref{flow1} and \eqref{msWIuntrunc}, appear at first sight to be in conflict with \eqref{equ:flowGamma} and \eqref{equ:sWiGamma} respectively.
In particular  factors of $\sqrt{\bar g}$ are apparently missing. This is because the $\sqrt{\bar g}$ factors are absorbed in a different definition of the inverse kernel. Indeed the inverse kernel \eqref{new-inverse} now satisfies
\be 
\left(\frac{\delta^{2}\Gamma_k}{\delta\varphi_x \delta\varphi_y}+r_{xy}\right) \Delta_{yz} =\delta_{xz}
\ee
without a $\sqrt{\bar g}(y)$  included in the integration over $y$.


Now that we have derived the flow equation and msWI written in a convenient notation, we are ready to prove that they are compatible. We begin by defining
\begin{equation}
\label{cal-W}
	\mathcal{W}_\omega\equiv\frac{\delta \Gamma_k}{\delta\chi_\omega}-\frac{\delta \Gamma_k}{\delta\varphi_\omega}
	-\frac{1}{2}\Delta_{xy}\frac{\delta r_{yx}}{\delta\chi_\omega} =0 \,.
\end{equation}
Taking the RG time derivative of $\mathcal{W_\omega}$ then gives
\begin{equation}
	\label{WIdot}
	\mathcal{\dot W}_\omega= \frac{\delta \dot\Gamma_k}{\delta\chi_\omega}
	-\frac{\delta \dot\Gamma_k}{\delta\varphi_\omega}+
	\frac{1}{2}\left[\Delta\left(\frac{\delta^{2}\dot{\Gamma}_k}{\delta\varphi \delta\varphi}
	+\dot r\right)\Delta\right]_{\!xy}\frac{\delta r_{yx}}{\delta\chi_\omega}
	-\frac{1}{2}\Delta_{xy}\frac{\delta \dot r_{yx}}{\delta\chi_\omega}
\end{equation}
and upon substituting the flow equation \eqref{flow1} into the right hand side, we have
\begin{align}\label{WIdot-2}
	\mathcal{\dot W}_{\omega}=&-\frac{1}{2}\Delta_{xz}
	\frac{\delta^{3}\Gamma_k}{\delta\varphi_{z}\delta\varphi_{z'}\delta\chi_{\omega}}\Delta_{z'y}\dot r_{yx}
	 + \frac{1}{2}\Delta_{xz}\frac{\delta^{3}\Gamma_k}{\delta\varphi_{z}
	 \varphi_{z'}\varphi_\omega}\Delta_{z'y}\dot{r}_{yx}
	 +\frac{1}{4}\Delta_{xz}\left(\frac{\delta^{2}}{\delta\varphi_{z}\delta\varphi_{z'}}\Delta_{uu'}\right)
	 \dot r_{u'u}\Delta_{z'y}\frac{\delta r_{yx}}{\delta\chi_{\omega}}\nonumber\\
	=&-\frac{1}{2}(\Delta \dot{r} \Delta)_{zz'}\frac{\delta^{2}}{\delta\varphi_{z'}\delta\varphi_{z}}
	 \left(\frac{\delta\Gamma}{\delta\chi_\omega}-\frac{\delta\Gamma}{\delta\varphi_\omega}\right)
	  +\frac{1}{4}\left(\frac{\delta^{2}}{\delta\varphi_{z}\delta\varphi_{z'}}\Delta_{uu'}\right)
	 \dot r_{u'u}\,\Delta_{z'y}\frac{\delta r_{yx}}{\delta\chi_{\omega}}\Delta_{xz} \,.
\end{align}
The first term in the last equality is in the form we want: a differential operator acting on (part of) $\mathcal{W_\omega}$. We now expand out the second term with the aim of also putting it into the desired form. For the sake of neatness let us define
\begin{equation}
		\Gamma_{x_1...x_n}\equiv \frac{\delta^n\Gamma_k}{\delta\varphi_{x_1}...\delta\varphi_{x_n}} \,.
\end{equation}
Expanding out the second term then gives
\begin{align}
\label{step1}
	\left(\frac{\delta^2}{\delta\varphi_z\delta\varphi_{z'}}\Delta_{uu'}\right)
	\dot r_{u'u}\,\Delta_{z'y}\frac{\delta r_{yx}}{\delta\chi_\omega}\Delta_{xz}
	=\Delta_{xz}& \bigg(\Delta_{uv}\Gamma_{zvs}\Delta_{sv'}\Gamma_{z'v's'}\Delta_{s'u'}
	+\Delta_{uv'}\Gamma_{v's'z'}\Delta_{s'v}\Gamma_{zvs}\Delta_{su'}\nonumber\\
	&\qquad-\Delta_{uv'}\Gamma_{v's'zz'}\Delta_{s'u'}\bigg)
	\dot r_{u'u}\Delta_{z'y}\frac{\delta r_{yx}}{\delta\chi_\omega} \,.
\end{align}
Upon exchanging factors of $\Delta$ and relabelling indices, we find
\begin{equation}
\label{step2}
	\left(\frac{\delta^2}{\delta\varphi_{z}\delta\varphi_{z'}}\Delta_{uu'}\right)
	\dot r_{u'u}\left(\Delta_{z'y}\frac{\delta r_{yx}}{\delta\chi_{\omega}}\Delta_{xz}\right)=
	(\Delta\dot r\Delta)_{s'v'}\frac{\delta^2}{\delta\varphi_{v'}\delta\varphi_{s'}} 
	\Delta_{xy}\frac{\delta r_{yx}}{\delta\chi_{\omega}} \,,
\end{equation}
which now has the structure we require. Thus we have shown that the RG time derivative of the msWI can be written as
\begin{equation}
	\mathcal{\dot W}_{\omega}=-\frac{1}{2}\text{tr}
	\left(\Delta \dot r\Delta\frac{\delta^2}{\delta\varphi\delta\varphi}\right)\mathcal{W}_{\omega} \,,
\end{equation}
\emph{i.e.} that it is proportional to the msWI itself. If $\Gamma_{k}$ satisfies $\mathcal{W_\omega}$ at some initial scale $k_0$, and satisfies the flow equation there, it thus follows without further restriction that $\mathcal{\dot W_\omega}|_{k_{0}}=0$ since it is proportional to $\mathcal{W_\omega}$. Thus the msWI is compatible with the flow equation. 
If $\Gamma_{k}$ continues to evolve according to the flow equation, 
it then follows that $\mathcal{W_\omega}$ and thus $\mathcal{\dot W_\omega}$ will be zero for all $k$. 

\subsection{Compatibility versus derivative expansion}\label{sec:exact-vs-derivatives}
\begin{figure}[ht]
\centering
\includegraphics[scale=0.23]{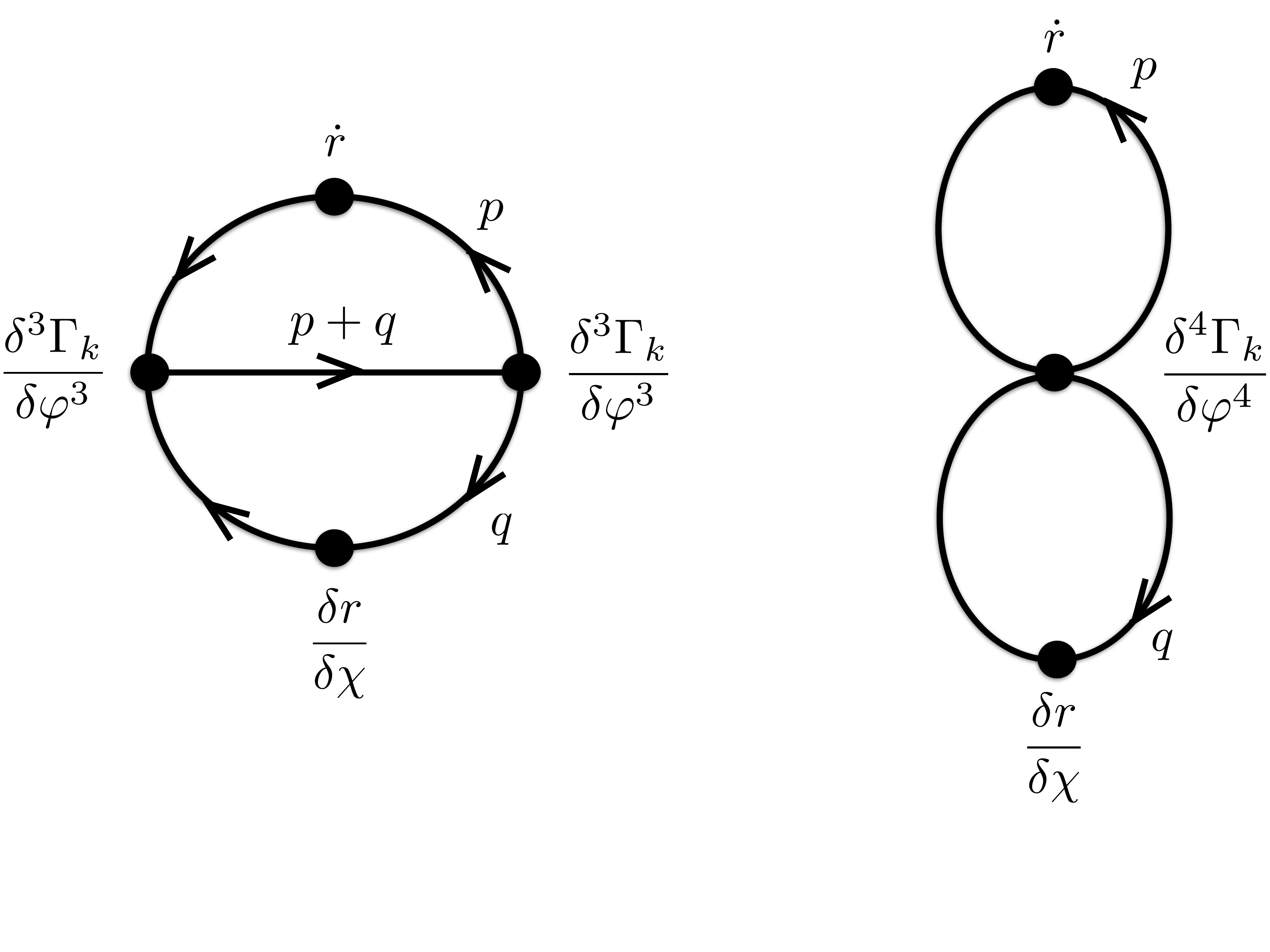}
\vskip-30pt
\caption{The two-loop diagrams in \eqref{step1}. Their symmetry immediately implies the identity \eqref{step2}. Momentum flow is indicated in the case where the fluctuation field $\varphi$ is then set to zero.}
\label{fig:two-loop}
\end{figure}
Recalling from \eqref{new-inverse} that $\Delta$ is an infrared regulated full propagator, we see from \eqref{step1} that the identity \eqref{step2} can be understood diagrammatically in terms of two-loop diagrams as sketched in fig. \ref{fig:two-loop}. 
The symmetry of these diagrams means that nothing changes if we exchange $\dot{r}\leftrightarrow\delta r/\delta\chi$. This exchange immediately leads to the identity \eqref{step2}.

This identity breaks down in general in the derivative expansion. If the Ward identity  is approximated by a derivative expansion, the full propagator in the one-loop term in \eqref{cal-W} is also expanded in a derivative expansion. This full propagator
 has loop momentum $q$ say, and is then expanded in powers of momenta carried by the external fluctuation field $\varphi(p)$, \ie by the external legs.
 The RG time derivative of the Ward identity yields the RG time derivative of such vertices, as can be seen from the $\delta^2\dot{\Gamma}_k/\delta\varphi^2$ term  in \eqref{WIdot}. This latter term has two internal legs given by the explicit functional derivatives, carrying the loop momentum $q$ and joining full internal propagators $\Delta$, and any number of external legs contained in the vertices of $\dot{\Gamma}_k$. Substituting the flow equation \eqref{flow1} then gives in particular the last term in eqn. \eqref{WIdot-2} in which two of these external legs are now joined to form a loop connected via $\dot{r}$. However it is momenta external to \emph{this new loop} which are Taylor expanded in the derivative expansion of the flow equation (see also \cite{Morris1999,Morris2001}). This is illustrated in the diagram displayed in fig. \ref{fig:two-loop}. In particular when the remaining external fluctuation field dependence is removed by setting $\varphi=0$, we have exactly the momentum dependence displayed in the figure. We see that a derivative expansion of the Ward identity involves Taylor expanding 
 in small $p$, while integrating over $q$. However a derivative expansion of the flow equation involves Taylor expanding 
 in small $q$, and integrating over $p$ instead. Thus the symmetry between the two loops is broken and the identity \eqref{step2} no longer follows.
 
 On the other hand we see that if $\dot{r}$ and $\delta r/\delta\chi$ have the same momentum dependence then the identity \eqref{step2} is restored because it is no longer possible to distinguish the two loops. Returning the placement of $\sqrt{\bar{g}}$ from \eqref{odd-r} to the integration measure, this in fact would give us the relation \eqref{same-p-dependence} that is necessary and sufficient for compatibility of the Ward identities within the derivative expansion, and which we will now derive directly within the derivative expansion.

\subsection{Compatibility at order derivative-squared}\label{sec:compatibility-at-d2}

We now proceed to calculate the flow of the msWI for the system truncated at $\mathcal{O}(\partial^2)$ as described in sec. \ref{sec:review},
and investigate directly under which circumstances it vanishes. Let us start by writing the flow equations and msWIs for both $V$ and $K$ in the following form so that we can study both cases simultaneously:
\begin{align}
	\label{flowA}
	\dot A(\varphi,\chi) &= \int_p B_p\dot R_p \,,\\
	\label{msWIA}
	\mathcal{W}^{(A)}&=\bar\partial A - \gamma A + \int_p B_p(\partial_\chi R_p + \gamma R_p)=0 \,,
\end{align}
where $A$ is either $V$ or $K/f$ such that $B_p$ is either $Q_p$ or $2 P_p$ respectively. Here we have also introduced the shorthand notation 
\be \label{gamma}
\int_p \equiv f(\chi)^{-\frac{d}{2}}\int dp \,p^{d-1}\,,\qquad \gamma \equiv \frac{d}{2}\partial_\chi \text{ln}f\,,\qquad{\rm and}\qquad \bar\partial \equiv \partial_\varphi - \partial_\chi\,.
\ee 
It will also be useful to have to hand the following relations:
\begin{align}
	\label{DV}
	\left( \bar \partial + \partial_t - \gamma \right) V &=
	\mathcal W^{(V)} + \int_q Q_q (\dot R_q - \partial_{\chi} R_q - \gamma R_q)\,,\\
	\label{DK}
	\left( \bar \partial + \partial_t - \gamma \right) \frac{K}{f} &=
	\mathcal W^{(K)} + 2 \int_q P_q (\dot R_q - \partial_{\chi} R_q - \gamma R_q)\,,\\
	\label{DQ}
	\left( \bar \partial + \partial_t + n \gamma \right) Q_p^n &=
	-n \, \, Q_p^{n+1} \int_q (\partial_{\varphi}^2 Q_q - 2 \, p^2 P_q)(\dot R_q - \partial_{\chi} R_q
	 - \gamma R_q)\nonumber\\ 
	&\qquad-n \, Q_p^{n+1} (\dot R_p - \partial_{\chi} R_p - \gamma R_p)
	-n \, Q_p^{n+1} (\partial_{\varphi}^2 \mathcal W^{(V)} - p^2 \mathcal W^{(K)}).
\end{align}
The first two relations are derived by subtracting the msWI from the flow equation for $V$ and $K/f$ respectively.   The last relation is then derived by using the first two relations above together with the definition of $Q_p$ given in \eqref{Q}.

We begin by taking the RG time derivative of \eqref{msWIA}. Substituting in the flow equation for $\dot A$, and remembering the power of $f(\chi)$ hidden in the integral over $p$, this gives
\begin{equation}
	\label{flowWA1}
	\dot{ \mathcal W } ^ {(A)} =
 	\int_p \dot R_p \left( \bar\partial + \partial_t + \gamma \right) B_p
 	- \int_p \dot B_p \left( \dot R_p - \partial_{\chi} R_p - \gamma R_p \right) .
\end{equation}
In order to proceed we have to assume a particular form of $B_p$ so that we can compute the result of the linear operators under the integral acting on it. A general term in $P_p$ takes the form
\begin{equation}
	\tilde B_p =
 	\left( \partial_{\varphi}^i V \right) ^ a
 	\left( \partial_{\varphi}^j \frac{K}{f} \right) ^ b
 	\left( \partial_{p^2}^k R_p \right) ^ c \left(p^2\right)^l Q_p^e \,,
\end{equation}
where $a,b,c,e,i,j,$ $k$ (not to be confused with the cutoff scale), and $l$ are non-negative integers. From the structure of the terms in $P_p$ one can read off the following sum rule for the exponents:
\begin{equation}
	a+b+c = e-1 \,.
\end{equation}
Notice that the case $B_p = Q_p$  for the potential is also included, since  $a=b=c=l=0$ and $e=1$ also satisfies the sum rule.
Taking the term under the first integral of \eqref{flowWA1}, we find
\begin{align}
\nonumber
	\left( \bar\partial + \partial_t + \gamma \right) \tilde B_p &=
	\bigg[a \left( \partial_{\varphi}^i V \right) ^ {-1} \partial_{\varphi}^i \left(  \bar\partial + \partial_t \right) V
	+ b \left( \partial_{\varphi}^j \frac{K}{f} \right) ^ {-1} \partial_{\varphi}^j
	 \left(  \bar\partial + \partial_t \right) \frac{K}{f} \\
	&\quad+ c \left( \partial_{p^2}^k R_p \right) ^ {-1} \partial_{p^2}^k \left( - \partial_{\chi} + \partial_t \right) R_p
	+ e \, Q_p^{-1} \left( \bar\partial + \partial_t \right) Q_p+ \gamma\bigg ] \tilde B_p \,.
\end{align}
Substituting equations \eqref{DV}--\eqref{DQ} into the above expression and using the sum rule, we obtain
\begin{align}
\label{DB}
	\nonumber
	\left( \bar\partial + \partial_t + \gamma \right) \tilde B_p &=
	\bigg[a \left( \partial_{\varphi}^i V \right) ^ {-1} \partial_{\varphi}^i \left(  \mathcal W^{(V)}
	 + \int_q Q_q \bar R_q \right)\\ \nonumber
	&\quad+ b \left( \partial_{\varphi}^j \frac{K}{f} \right) ^ {-1} \partial_{\varphi}^j \left(  \mathcal W^{(K)}
	+ 2 \int_q P_q \bar R_q \right)
	+ c \left( \partial_{p^2}^k R_p \right) ^ {-1} \partial_{p^2}^k \bar R_p \\
	&\quad- e\, Q_p \int_q \left( \partial_{\varphi}^2 Q_q - 2 p^2 P_q \right) \bar R_q
	- e\, Q_p \bar R_p
	- e\, Q_p \left( \partial_{\varphi}^2 \mathcal W^{(V)} - p^2 \mathcal W^{(K)} \right)
	 \bar R_q\bigg] \tilde B_p \,.
\end{align}
where we have introduced the shorthand notation 
\be 
\bar R_p = \dot R_p - \partial_\chi R_p - \gamma R_p\,.
\ee 
Turning our attention now to the second integral of \eqref{flowWA1} we take the RG time derivative of $\tilde B_p$ and again substitute in the flow equations for $V$ and $K/f$. This gives
\begin{align}
\label{Bdot}
\nonumber
	\dot{ \tilde B }_p &=
	\bigg[	a \left( \partial_{\varphi}^i V \right) ^ {-1} \partial_{\varphi}^i \int_q Q_q \dot R_q
	+ b \left( \partial_{\varphi}^j \frac{K}{f} \right) ^ {-1} \partial_{\varphi}^j \int_q 2 P_q \dot R_q \\
	&\quad+ c \left( \partial_{p^2}^k R_p \right) ^ {-1} \partial_{p^2}^k \dot R_p
	- e \, Q_p \int_q  \left( \partial_{\varphi}^2 Q_q - 2 p^2 P_q \right) \dot R_q
	- e \, Q_p  \dot R_q\bigg] \tilde B_p  \,.
\end{align}
Inserting \eqref{DB} and \eqref{Bdot} into \eqref{flowWA1} we obtain
\begin{align}
\nonumber
 	\mathcal{\dot W}^{(A)} &= \sum_{\tilde B_p} \Bigg\{a \int_{p,q} 
 	\tilde B_p ( \partial_{\varphi}^i V ) ^ {-1} \partial_{\varphi}^i
	\bigg( \dot R_p \mathcal{W}^{(V)} +Q_q [ \dot R , \partial_{\chi} R + \gamma R ]_{qp} \bigg) \\ \label{flowWA}
	&\quad+ b \int_{p,q} \tilde B_p \left( \partial_{\varphi}^j \frac{K}{f} \right) ^ {-1} \partial_{\varphi}^j
	\bigg( \dot R_p \mathcal{W}^{(K)} + 2 P_q [ \dot R , \partial_{\chi} R + \gamma R ]_{qp} \bigg) \\\nonumber
	&\quad+ c \int_p \tilde B_p \left( \partial_{p^2}^k R_p \right) ^ {-1}
	\left( \left(\partial_\chi R_p + \gamma R_p\right) \partial_{p^2}^k \dot R_p - \dot R_p \partial_{p^2}^k \left(\partial_\chi R_p + \gamma R_p\right) \right) \\ \nonumber
	&\quad- e \int_p \tilde B_p  Q_p  \dot R_p \bigg( \partial^2_{\varphi} \mathcal{W}^{(V)}
	 - p^2  \mathcal{W}^{(K)} \bigg)- e \int_{p,q}  \tilde B_p Q_p \bigg(  \partial_{\varphi}^2 Q_q 
	 - 2 p^2 P_q \bigg) [ \dot R , \partial_{\chi} R + \gamma R ]_{qp}\Bigg\}\,,
\end{align}
where we have introduced the commutator-like construct $[A,B]_{qp} = A_q B_p - B_qA_p$.

When $A = V$ the above expression simplifies considerably to
\begin{equation}
\label{flowWV}
	 \mathcal{\dot W}^{(V)}=-  \int_p  Q^2_p  \dot R_p\, \bigg( \partial^2_{\varphi} \mathcal{W}^{(V)}
	  - p^2  \mathcal{W}^{(K)} \bigg) - \int_{p,q}   Q^2_p\, \bigg(  \partial_{\varphi}^2 Q_q
	  - 2 p^2 P_q \bigg) [ \dot R , \partial_{\chi} R + \gamma R ]_{qp} \,,
\end{equation}
which we see contains only terms that contain either the Ward identities or the `commutator' $[ \dot R , \partial_{\chi} R + \gamma R ]_{qp}$. On the other hand for the flow of the $K/f$ msWI, the terms do not collect, so that it remains separately dependent on the individual $\tilde B_p$. However each term either contains the Ward identities themselves, the `commutator'  $[ \dot R , \partial_{\chi} R + \gamma R ]_{qp}$, or the additional commutator-like structures:
\be
\label{k-comm}
 \left(\partial_\chi R_p + \gamma R_p\right) \partial_{p^2}^k \dot R_p - \dot R_p \partial_{p^2}^k \left(\partial_\chi R_p + \gamma R_p\right)\,.
 \ee
These appear in the third line of \eqref{flowWA}, and the integer $k$ takes values $1$ and $2$. For a general cutoff $R_p$, these two additional commutator terms neither vanish nor combine with other terms of the flow.

If $[ \dot R , \partial_{\chi} R + \gamma R ]_{qp}$ vanishes, the flow \eqref{flowWV} of the V msWI is automatically satisfied providing that both the $K$ and $V$ msWI are also satisfied. In this case we have by rearrangement that
\be
\left( \partial_\chi R_p + \gamma R_p \right)/\dot R_p = \left( \partial_\chi R_q + \gamma R_q \right)/\dot R_q\,,
\ee
which means that the ratio is independent of momentum. Equivalently
\be
\label{same-p-dependence}
   \partial_\chi R_p + \gamma R_p = F(\chi,t) \,\dot R_p \,,
\ee
where $F$ can be a function of $\chi$ and $t$ but not of $p$. 
However it is straightforward to see that \eqref{same-p-dependence} also forces the additional commutators \eqref{k-comm} to vanish.

We have therefore shown that all the commutator-like terms vanish if and only if $\dot R_p$ and $\partial_{\chi} R_p + \gamma R_p$ have the same dependence on $p$, with the consequence  that both the $\mathcal{\dot W}^{(A)}$ vanish, if the Ward identities $\mathcal{W}^{(A)}$ themselves vanish. Since for general choices of the functions, the vanishing of the `commutators' is surely necessary to achieve $\mathcal{\dot W}^{(A)}=0$ without further restriction, we have thus shown that the condition \eqref{same-p-dependence} is necessary and sufficient to ensure compatibility, as defined at the beginning of this section.

\subsection{Incompatibility implies no solutions}\label{sec:incompatibility}

However even if the commutators do not vanish, and thus the Ward identities are incompatible with the flow equations, {\it a priori} there could still be a non-empty restricted set of solutions that both satisfy the flow equations and Ward identities. In this case the equations are satisfied not by the vanishing of the commutators themselves, but by the fact that for the given solutions the sum of all these terms vanish after performing the integration over momenta. 
Therefore, as well as obeying the flow equations and the msWIs $\mathcal{W}^{(A)}=0$, the solutions must also separately obey two further conditions, namely the vanishing of the right hand sides of \eqref{flowWA}. In the language of Dirac's classification of constraints \cite{Dirac-primary,Dirac1950}, the $\mathcal{W}^{(A)}=0$ provide the primary constraints. We have shown that if the `commutators' do not vanish, then the solutions are subject also to non-trivial secondary constraints $\mathcal{\dot W}^{(A)}=0$. Given the involved form of $\mathcal{\dot W}^{(K)}$ in particular,  we can be sure that the procedure does not close and that actually there is then an infinite tower of secondary constraints, $\partial^n_t \,\mathcal{W}^{(A)}=0,\;\forall \,n>0$,  all of which must be satisfied. It would therefore seem inevitable that there are in fact no non-trivial solutions in this case. We will confirm this by example in sec. \ref{sec:incompatible-no-solns}. We conclude that \emph{the vanishing of the `commutators', and hence condition \eqref{same-p-dependence}, is both necessary and sufficient for there to be any solutions to the flows and Ward identities in the derivative expansion approximation outlined in sec. \ref{sec:review}}.

The condition \eqref{same-p-dependence} was already used in ref. \cite{Dietz:2015owa}, where however it was introduced as a mathematical trick to help solve the coupled system of flow equations and msWI. As we recall below, it implies either that $\eta=0$ or $R_p$ is of power-law form. We now see that the requirement for $\dot R_p$ and $\partial_{\chi} R_p + \gamma R_p$ to have the same dependence on $p$, goes much deeper: the flow equations \eqref{flowV} and \eqref{flowK}, and the Ward identities \eqref{msWIV} and \eqref{msWIK}, are incompatible without this constraint, and incompatibility forces there to be no solutions to the combined system.


\subsection{Required form of the cutoff profile}\label{sec:required-cutoff}

Note that $R_p$ must take a form that respects the scaling dimensions. Introducing dimensionless variables for use in the next section and later, we can make these scaling dimensions explicit by employing the RG scale $k$. 
We denote the new dimensionless quantities with a bar. We have
\begin{align}
\label{dim-vars}
	\varphi = k^{\eta/2}\bar\varphi, \qquad\chi = k^{\eta/2}\bar\chi, \qquad f(\chi) = k^{d_f}\bar f(\chi),\nonumber\\
	V(\varphi,\chi) = k^{d_V}\bar V(\bar\varphi,\bar\chi), \qquad K(\varphi, \chi) = k^{d_R- 2 + d_f} \bar  K(\bar\varphi,\bar\chi),
\end{align}
where 
\be 
\label{dimensions}
d_V = d(1-d_f/2)\qquad{\rm and}\qquad d_R = d_V - \eta\,,
\ee 
and thus from \eqref{cutoff-action} and \eqref{conformal-reduction}, we have by dimensions that $R_p$ must take the form
\begin{equation}
\label{equ:cutoff}
	R(p^2/f)= - k^{d_R} \,r\left(\frac{p^2}{k^{2-d_f}f}\right) = - k^{d_R} \,r(\hat p^2) \,,
\end{equation}
where $r$ is a dimensionless cutoff profile of a dimensionless argument,\footnote{The minus sign in \eqref{equ:cutoff} is necessary to work with the wrong sign kinetic term in \eqref{trunc} \cite{Dietz:2015owa}.} and we have introduced the dimensionless momentum magnitude $\hat p = p\sqrt{k^{d_f-2}/f}$. 

If $\dot R_p$ and $\partial_{\chi} R_p + \gamma R_p$ have the same dependence on $p$, \ie satisfy \eqref{same-p-dependence}, then either $\eta=0$ or $R_p$ is of power-law form \cite{Dietz:2015owa}. To see this, note that from \eqref{equ:cutoff} and \eqref{gamma} we have
\be
\label{R-scaling-relation}
\gamma {\dot R}_p = d_V \left[\partial_\chi R_p+\gamma R_p\right] -\eta \gamma R_p\,.
\ee
Thus (choosing $F=\gamma/d_V$) we see that \eqref{same-p-dependence} is satisfied if $\eta=0$, without further restriction on $R$. However if $\eta\ne0$, then \eqref{R-scaling-relation} together with \eqref{same-p-dependence} implies
\be 
f \frac{\partial R_p}{\partial f} = \frac{d}{2}\left( \frac{\eta F}{ d_VF- \gamma} - 1\right) R_p\,,
\ee
and thus from \eqref{equ:cutoff}
\be 
\hat{p} \frac{d}{d\hat{p}} r(\hat{p}^2) = -d\left( \frac{\eta F}{ d_VF- \gamma} - 1\right) r(\hat{p}^2)\,.
\ee
Since the term in brackets does not depend on $p$, we see that this is only possible if in fact the term in brackets is a constant. Setting this constant to be $2n/d$ for some constant $n$, we thus also deduce that $r\propto \hat{p}^{-2n}$.

An example of a cutoff that does not satisfy \eqref{same-p-dependence} if $\eta\ne0$, and thus leads to incompatible msWIs in this case, is the optimised cutoff \cite{opt1,opt3}:
\be 
\label{optimised}
r(\hat{p}^2) = (1-\hat{p}^2)\theta(1-\hat{p}^2)\,.
\ee
It is straight-forward to confirm that this does not satisfy \eqref{same-p-dependence} if $\eta\ne0$. Using \eqref{equ:cutoff} and \eqref{R-scaling-relation} we find
\be 
\dot{R}_p \propto d_V \left[ \frac{2}{d}\,\theta(1-\hat{p}^2) + (1-\hat{p}^2)\theta(1-\hat{p}^2)\right] - \eta\, (1-\hat{p}^2)\theta(1-\hat{p}^2)\,.
\ee
In order for \eqref{optimised} to satisfy \eqref{same-p-dependence}, the right hand side must be proportional to $\partial_\chi R_p+\gamma R_p$ \ie to the term in square brackets. This is only true if $\eta=0$.

\section{LPA equations}

We will now use the Local Potential Approximation to further investigate the restriction imposed by the msWI on the RG flow equation, in terms of general solutions and also on the existence of $k$-fixed points (\ie RG fixed points with respect to variations in $k$). We start with a very clear example where the msWI forbids the existence of $k$-fixed points. 

Then using the concrete example of the optimised cutoff we show explicitly that compatibility forces $\eta=0$ for non-power-law cutoffs. Setting $\eta=0$ we will see that background independent variables exist, in other words they exist whenever the msWI is compatible with the flow. We will also see that such $\hat{k}$-fixed points coincide with the $k$-fixed points. The background independent variables allow us to solve for the fixed points explicitly, uncovering a line of fixed points, consistent with the findings for power-law cutoff \cite{Dietz2016}.

\subsection{Demonstration of background independence forbidding fixed points in general}\label{sec:forbids}

We use the change to dimensionless variables \eqref{dim-vars} and \eqref{equ:cutoff}. In the LPA we discard the flow and Ward identity for $K$, and set $\bar K=1$. The result, for general cutoff profile $r(\hat{p}^2)$, is:
\begin{align}
\label{flow}
\partial_t \bar V + d_V \bar V - \frac{\eta}{2} \, \bar\varphi \, \frac{\partial \bar V}{\partial \bar\varphi} - \frac{\eta}{2} \, \bar\chi \, \frac{\partial \bar V}{\partial \bar\chi} &=
\int_0^{\infty} d\hat p \, \hat p^{d-1} \, \frac{d_R\, r - \frac{d_V}{d} \, \hat p \, r'}{\hat p^2 + r - \partial^2_{\bar\varphi}\bar V}\,,\\
\label{msWI}
\frac{\partial \bar V}{\partial \bar\chi} - \frac{\partial \bar V}{\partial \bar\varphi} + \bar \gamma \, \bar V &= \bar \gamma
\int_0^{\infty} d\hat p \, \hat p^{d-1} \, \frac{r - \frac{1}{d} \, \hat p \, r'}{\hat p^2 + r - \partial^2_{\bar\varphi}\bar V} \,,
\end{align}
where $r'$ means $dr(\hat{p}^2)/d\hat{p}$ and from the change to dimensionless variables we find:
\be 
\label{fbar-general}
\bar{\gamma} = \frac{d}{2}\frac{\partial}{\partial\bc}\ln\bar{f}\left({\rm e}^{\eta t/2}\mu^{\eta/2}\bc\right)\,.
\ee
Note that since $f$ cannot depend on $t$ (see the discussion in sec. \ref{sec:review}), once we go to dimensionless (\ie scaled) variables, $\bar{f}$ is in general forced to depend on $t$ if $\chi$ has non-vanishing scaling dimension $\eta$.
At the ($k$-)fixed point we must have $\partial_t \bar V = 0$. We see at once why fixed points are generically forbidden by the msWI: the fixed point potential $\bar{V}$ would have to be independent of $t$, but through \eqref{msWI} and \eqref{fbar-general} this is impossible in general since $\bV$ is forced to be dependent on explicit $t$-dependence in $\bar{f}$ through the Ward identity. This is true even in the case of power-law cutoff profile\footnote{And indeed this issue was highlighted, but in a different way in ref. \cite{Dietz:2015owa}.} which as we have seen allows \eqref{msWI} to be compatible with the flow \eqref{flow}.

At first sight an escape from this problem is simply to set $f$ to be power law. Indeed setting $f\propto\chi^{\rho}$
for some constant $\rho$,  \eqref{fbar-general} implies
\be 
\bg = \frac{d}{2} \frac{\rho}{\bc}\,,
\label{powlaw-gamma}
\ee 
and thus \eqref{msWI} no longer has explicit $t$ dependence. 
Recall that for power-law cutoff profiles $r$, it was indeed found that $k$-fixed points for $\bV$ 
are allowed if $f$ is chosen of power law form \cite{Dietz:2015owa}.\footnote{This is true also for $\bar{K}$. However if the dimensions of $f$ and $\chi$
do not match up, 
these fixed points do not agree with the background independent $\hat{k}$-fixed points and furthermore  the effective action $\Gamma_k$ still runs with $k$ \cite{Dietz:2015owa}.} 
However we have seen in sec. \ref{sec:required-cutoff} that any other cutoff profile does not allow the Ward identity to be compatible with the flow unless $\eta=0$. We argued in sec. \ref{sec:incompatibility} that incompatibility overconstrains the equations leading to no solutions. In the next subsection, sec. \ref{sec:incompatible-no-solns}, we will confirm this explicitly, choosing as a concrete example the optimised cutoff profile and space-time dimension $d=4$.

On the other hand, if we set $\eta=0$ then the msWI \eqref{msWI} is compatible with the flow \eqref{flow}, for any parametrisation $f$. Apparently $k$-fixed points are also now allowed without further restriction, since again \eqref{fbar-general} loses its explicit $t$ dependence.  Opting once more for optimised cutoff profile and $d=4$, we will see in sec. \ref{sec:works} that indeed they are allowed and furthermore they coincide with fixed points in a background independent description that we also uncover.

\subsection{Confirmation of no solutions if the msWI is incompatible with the flow}\label{sec:incompatible-no-solns}

Specialising to optimised cutoff and (for simplicity) the most interesting case of spacetime dimension $d=4$, the equations read
\begin{align}
\label{flowComplete_opt}
\partial_t\bar{V}+
d_V \bar V - \frac{\eta}{2} \bar \varphi \, \partial_{\bar\varphi}\bar V -\frac{\eta}{2} \bar \chi \, \partial_{\bar\chi}\bar V &= \left(  \frac{d_R}{6} + \frac{\eta}{12} \right) \frac{1}{1 -  \partial^2_{\bar\varphi}\bar V},\\
\label{msWI_opt}
\partial_{\bar\chi}\bar V - \partial_{\bar\varphi}\bar V + \bar{\gamma} \bar V &= \frac{\bar{\gamma}}{6} \frac{1}{1 - \partial^2_{\bar\varphi}\bar V}\,.
\end{align}
Choosing power law $f$ and thus \eqref{powlaw-gamma} there is no explicit $t$ dependence and apparently these equations can work together. 
Combining them 
by eliminating their right hand sides, we get
\be
\label{pde1}
2\partial_t\bar{V}+\eta\bar{V}-\left(\eta\bar\varphi-\alpha\bc\right)\partial_{\bar{\varphi}}\bar{V}-(\eta+\alpha)\bar\chi\partial_{\bar{\chi}} \bar{V}=0\,,
\ee
where we have introduced the constant $\alpha = (d_R+\eta/2)/\rho$.
This equation can be solved by the method of characteristics (see \eg the appendix in ref. \cite{Dietz:2015owa}). Parametrising the characteristic curves with $t$, they are generated by the following equations:
\be 
\frac{d\bV}{dt} = -\frac{\eta}{2}\bV\,,\quad\frac{d\bc}{dt}=-\frac{\alpha+\eta}{2}\bc\,,\quad \frac{d\bp}{dt}=\frac{\alpha\bc-\eta\bp}{2}\,.
\ee
Solving the second equation before the third, it is straightforward to find the curves:
\be 
\bV = \hV {\rm e}^{-\eta t/2}\,,\quad \bc = \hc\, {\rm e}^{-(\eta+\alpha)t/2}\,,\quad \bp+\bc = \hp\, {\rm e}^{-\eta t/2}\,,
\ee
in terms of initial data $\hV,\hp,\hc$.
Thus the solution to \eqref{pde1} can be written as
\be 
\bV ={\rm e}^{-\eta t/2}\, \hV(\hp,\hc) ={\rm e}^{-\eta t/2} \,\hV\left({\rm e}^{\eta t/2}[\bp+\bc],{\rm e}^{(\eta+\alpha)t/2}\bc\right)\,,
\ee
as can be verified directly. Plugging this into either \eqref{flowComplete_opt} or \eqref{msWI_opt} gives the same equation, which in terms of the hatted variables reads
\be 
\hc \partial_{\hc}\hV+2\rho\hV = \frac{\rho}{3}\frac{1}{{\rm e}^{-\frac{\eta}{2} t} \,-\partial^2_{\hp}\hV}\,.
\ee
Since $\hV(\hp,\hc)$ is independent of $t$, we see there are no solutions unless $\eta=0$. 
We saw in sec. \ref{sec:required-cutoff} that this was also the necessary and sufficient condition for compatibility in this case. 

\subsection{Background independence at vanishing anomalous dimension}\label{sec:works}
We now set $\eta=0$. As recalled in sec. \ref{sec:required-cutoff}, the msWI is now compatible with the flow, and furthermore  from \eqref{fbar-general} the explicit $t$ dependence has dropped out.
For power-law cutoff profiles we found that $k$-fixed points exist and coincide with background independent $\hk$-fixed points for any form of $f$ with any dimension $d_f$ \cite{Dietz:2015owa}. We will see that for non-power law cutoff that the same is true. (Again we choose optimised cutoff and $d=4$ as an explicit example.) We will uncover consistent background independent variables for which the full line of fixed points is visible \cite{Dietz2016}. 

Since $\eta=0$, in the equations \eqref{flowComplete_opt} and \eqref{msWI_opt}, we also have $d_R=d_V=2(2-d_f)$ and $\bar{\gamma}=2\partial_{\bc}\ln\bar{f}(\bc)$. Note that from \eqref{equ:cutoff}, $d_f=2$ is excluded otherwise the IR cutoff no longer depends on $k$. Also note that since $\eta=0$ we can drop the bars on $\chi$ and $\vp$.
Combining the equations into a linear partial differential equation we get
\be 
\label{pde2}
\partial_t\bV +\frac{2-d_f}{\partial_{\chi}\ln\bar{f}}\left(\partial_{\vp}\bV-\partial_{\chi}\bV\right) =0\,,
\ee
whose characteristic curves satisfy
\be 
\label{eta0curves}
\frac{d\chi}{dt}=\frac{d_f-2}{\partial_{\chi}\ln\bar{f}}\,,\quad\frac{d\vp}{dt}=\frac{2-d_f}{\partial_{\chi}\ln\bar{f}}\,,\quad \frac{d\bV}{dt}=0\,.
\ee
Solving the first equation gives:
\be 
\label{hatt}
\hatt = t+\frac{\ln\bar{f}}{2-d_f}\,,
\ee
where the integration constant $\hatt$ is thus the background independent definition of RG time (see the appendix to ref. \cite{Dietz:2015owa}). Exponentiating,
\be 
\hk = k \left\{\bar{f}(\chi)\right\}^{\frac{1}{2-d_f}} = k^{2\frac{1-d_f}{2-d_f}} \left\{f(\chi)\right\}^{\frac{1}{2-d_f}}\,,
\ee
where the second equality follows from \eqref{dim-vars}. The sum of the first two equations in \eqref{eta0curves} tells us that $\phi=\vp+\chi$ is an integration constant for the characteristics, and finally the last equation says that $\bV$ is also constant for characteristics. Thus we learn that the change to background independent variables is achieved by writing
\be 
\label{background-independent}
\bV = \hV(\phi,\hatt\,)\,.
\ee
It is straightforward to verify that this solves \eqref{pde2}. Substituting into either \eqref{flowComplete_opt} or \eqref{msWI_opt} gives the same flow equation:
\be 
\label{bi-flow}
\partial_{\hatt}\hV +d_V\hV = \frac{d_V}{6}\frac{1}{1-\partial^2_\phi\hV^{\vphantom{H^H}}}\,,
\ee
which is indeed now background independent, \ie independent of $\chi$, and indeed independent of parametrisation $f$. There remains a dependence on the dimension of $f$ through $d_V = 2(2-d_f)$ although this disappears for $\hk$-fixed points, and can be removed entirely by a rescaling $\hatt\mapsto \hatt\, d_V$ which however changes the dimension of $\hat{k}$ to $d_V$. 

We also see from \eqref{hatt} and \eqref{background-independent} that
\be 
\partial_t \bV = \partial_{\hatt} \hV\,,
\ee
and thus fixed points in $k$ coincide with the background independent fixed points. 

Finally, the fixed points are readily found from \eqref{bi-flow} similarly to refs. \cite{Dietz2016,Morris:1994jc} by recognising that 
\be 
\frac{d^2\hV}{d\phi^2} = 1-\frac{1}{6\hV}
\ee
is equivalent to Newton's equation for acceleration with respect to `time' $\phi$ of a particle of unit mass at `position' $\hV$ in a potential $U=-\hV+ (\ln\hV)/6$. In this way it can be verified that there is a line of fixed points ending at the Gaussian fixed point, which is here $\hV=1/6$, in agreement with the findings for power-law cutoff in \cite{Dietz2016}.

\section{Polynomial truncations}
\label{sec:truncations} 

The analysis  so far has used properties of conformally truncated gravity and the derivative expansion approximation method. In order to gain insight about what might happen at the non-perturbative level, and in full quantum gravity, we will consider how the issues would become visible in polynomial truncations. 

The generic case treated in sec. \ref{sec:forbids} will be just as clear in the sense that truncations of the Ward identity will still force the effective potential (effective action in general) to be $t$ dependent if the dimensionless parametrisation \eqref{fbar-general} is similarly forced to be $t$ dependent. In general therefore, if the way the metric is parametrised forces the parametrisation to become $t$ dependent, we can expect that background independence excludes the possibility of fixed points, at least with respect to $t$.

Consider next the situation treated in sec. \ref{sec:incompatible-no-solns}. Expanding the dimensionless potential and the equations in a double power series in the fluctuation and the background field, we write:
\begin{equation}
\label{pot-expand}
\bar V(\bar\varphi,\bar\chi) = \sum_{n,m=0}^{\infty} a_{nm} \bar\varphi^n \bar\chi^m\,.
\end{equation}
Substituting \eqref{powlaw-gamma} into \eqref{msWI_opt} and multiplying through by $\bc$, we can read off from this and \eqref{flowComplete_opt} the zeroth level equations:
\be 
d_V a_{00} = \left(  \frac{d_R}{6} + \frac{\eta}{12} \right) \frac{1}{1 -  2a_{20}}\,,\qquad 2\rho\, a_{00} = \frac{\rho}{3} \frac{1}{1 -  2a_{20}}\,.
\ee
Since $\rho$ cannot vanish and $a_{20}$ cannot diverge, combining these equations gives $d_V = d_R+\eta/2$ which from \eqref{dimensions} implies $\eta=0$. Thus we recover already from the zeroth order level that fixed points are excluded unless $\eta=0$. (Of course the real reason, namely that the equations are incompatible, and the full consequence that there are no $t$-dependent solutions either, is maybe not so easy to see this way.)

\subsection{Counting argument}\label{sec:counting}


We already argued in the Introduction, that generically the coefficients become overconstained if we consider a sufficiently high truncation. We now proceed to make a careful count and estimate the level at which this happens.

We concentrate on fixed point solutions to the LPA system \eqref{flow}, \eqref{msWI} and \eqref{fbar-general} where either $\eta=0$ or we choose power-law $f$, so that explicit $t$ dependence does not already rule out such solutions. We introduce the short-hand notation $\bar V ^{(n,m)}=\partial_{\bar \varphi}^n \partial_{\bar \chi}^m \bar V(\bp,\bc)$. 
To obtain the system at order $r$ we have to plug the expansion of the potential \eqref{pot-expand}
into both the fixed point equation and msWI, act on them with operators $\frac{\partial^{i+j}}{\partial \bar\varphi^i \, \partial \bar\chi^j}$ such that $i+j=r$, before finally setting the fields to zero. In particular, for any fixed value $r_{\star}$ we have $2 \, (r_{\star}+1)$ equations
and hence up to order $r$ there are
\begin{equation}
\label{number_eqns}
n_{\text{eqn}}(r) = \sum_{i=0}^{r} 2\,(i+1) = r^2 + 3r + 2 
\end{equation}
equations. 

To count the coefficients appearing in these $n_{\text{eqn}}(r)$ equations let us start with the left hand sides. First note that 
\begin{equation}
\label{proportionality}
\bar V^{(i,j)} \bigg|_{\bar\varphi = \bar\chi = 0}
\propto a_{ij} \,.
\end{equation}
That is, for any fixed pair $(i,j)$ the left hand side of (\ref{msWI}) will contain the coefficients $a_{ij}$, $a_{i+1,j}$ and $a_{i,j+1}$, whereas the left hand side of (\ref{flow}) will only contain $a_{ij}$. Up to some fixed order $r$ there will be thus coefficients $a_{ij}$ where $i$ and $j$ run from $0$ to $r+1$ and $i+j \leqslant r+1$
\begin{figure}
\centering
\begin{tikzpicture}[>=stealth]
\draw (0,0) -- (1,0) -- (1,-1) -- (0,-1) -- cycle;
\draw (1,0) -- (2,0) -- (2,-1) -- (1,-1) -- cycle;
\draw (2,0) -- (3,0) -- (3,-1) -- (2,-1) -- cycle;
\draw (3,0) -- (4,0) -- (4,-1) -- (3,-1) -- cycle;

\draw (0,-1) -- (1,-1) -- (1,-2) -- (0,-2) -- cycle;
\draw (1,-1) -- (2,-1) -- (2,-2) -- (1,-2) -- cycle;
\draw (2,-1) -- (3,-1) -- (3,-2) -- (2,-2) -- cycle;

\draw (0,-2) -- (1,-2) -- (1,-3) -- (0,-3) -- cycle;
\draw (1,-2) -- (2,-2) -- (2,-3) -- (1,-3) -- cycle;

\draw (0,-3) -- (1,-3) -- (1,-4) -- (0,-4) -- cycle;

\draw (0.5,0.5) node {0}  (1.5,0.5) node {1}  (2.5,0.5) node {$\cdots$} (3.5,0.5) node {$r+1$};
\draw (-0.5,-0.5) node {0}  (-0.5,-1.5) node {1}  (-0.5,-2.5) node {$\vdots$} (-0.5,-3.5) node {$r+1$};

\draw (0.5,-0.5) node {$a_{00}$}  (1.5,-0.5) node {$a_{01}$}
(2.5,-0.5) node {$\cdots$}  (3.5,-0.5) node {$a_{0,r+1}$};
\draw (0.5,-1.5) node {$a_{10}$}  (1.5,-1.5) node {$\ddots$};
\draw (0.5,-2.5) node {$\vdots$};
\draw (0.5,-3.5) node {$a_{r+1,0}$};

\draw (-0.25,0.5) node {$j$} (-0.5,0.25) node {$i$};
\draw (-0.15,0.15) -- (-0.55,0.55);
\end{tikzpicture}
\caption{Coefficients of the potential appearing on the left sides of the equations.}
\label{diag_lhs}
\end{figure}
\begin{equation}
\label{coeff_lhs}
\bigg\lbrace a_{00}, a_{01}, \dots , a_{0,r+1}, a_{10}, \dots , a_{1,r}, \dots , a_{2,r-1}, \dots, a_{r+1,0}   \bigg\rbrace \,,
\end{equation}
 (cf. figure \ref{diag_lhs}). This adds up to the following number of coefficients
\begin{equation}
n_{\text{lhs}}(r) = \sum_{i=1}^{r+2} \, i = \frac{1}{2} \, r^2 + \frac{5}{2} \, r + 3 \,.
\end{equation}

Including the coefficients of the right hand side, we have to be careful not to double count any coefficients that have already been taken account of on the left hand sides. Let us suppose we have fixed the cutoff and let us assume that for the moment $\bar{\gamma} =$ const. Then all additional coefficients on the right hand side come from the expansion of the propagator
\begin{equation}
\label{expansion_propagator}
\frac{\partial^{i+j}}{\partial \bar\varphi^i \, \partial \bar\chi^j} \, \left( \frac{1}{1 - \bar V^{(2,0)}} \right) \bigg|_{\bar\varphi = \bar\chi = 0} = 
\frac{\partial^{j}}{\partial \bar\chi^j} \left[ \frac{\partial^{i-1}}{\partial \bar\varphi^{i-1}} \left( \frac{\bar V^{(3,0)}}{(1 - \bar V^{(2,0)})^2} \right) \right] \bigg|_{\bar\varphi = \bar\chi = 0}\, .
\end{equation}
Since we can always arrange the $\bar\varphi$--derivatives to act first, the expression in the square brackets will involve terms $\bar V^{(2,0)} \cdots \, \bar V^{(i+2,0)}$. Using (\ref{proportionality}), we see that the expression given in (\ref{expansion_propagator}) will then include terms
\begin{equation}
\bigg\lbrace a_{20}, a_{21}, \dots , a_{2i}, a_{30}, \dots , a_{3i}, \dots, a_{i+2,0}, \dots, a_{i+2,j}   \bigg\rbrace \,.
\end{equation}
Up to any fixed order $r$, $i$ and $j$ can take values between $0$ and $r$ such that $i+j = r$, and in total we will have the following coefficients on the right hand sides 
\begin{figure}
\centering
\begin{tikzpicture}[>=stealth]
\draw (0,0) -- (1,0) -- (1,-1) -- (0,-1) -- cycle;
\draw (1,0) -- (2,0) -- (2,-1) -- (1,-1) -- cycle;
\draw (2,0) -- (3,0) -- (3,-1) -- (2,-1) -- cycle;
\draw (3,0) -- (4,0) -- (4,-1) -- (3,-1) -- cycle;

\draw (0,-1) -- (1,-1) -- (1,-2) -- (0,-2) -- cycle;
\draw (1,-1) -- (2,-1) -- (2,-2) -- (1,-2) -- cycle;
\draw (2,-1) -- (3,-1) -- (3,-2) -- (2,-2) -- cycle;

\draw (0,-2) -- (1,-2) -- (1,-3) -- (0,-3) -- cycle;
\draw (1,-2) -- (2,-2) -- (2,-3) -- (1,-3) -- cycle;

\draw (0,-3) -- (1,-3) -- (1,-4) -- (0,-4) -- cycle;

\draw (0.5,0.5) node {0}  (1.5,0.5) node {1}  (2.5,0.5) node {$\cdots$} (3.5,0.5) node {$r$};
\draw (-0.5,-0.5) node {2}  (-0.5,-1.5) node {3}  (-0.5,-2.5) node {$\vdots$} (-0.5,-3.5) node {$r+2$};

\draw (0.5,-0.5) node {$a_{20}$}  (1.5,-0.5) node {$a_{21}$}
(2.5,-0.5) node {$\cdots$}  (3.5,-0.5) node {$a_{2,r}$};
\draw (0.5,-1.5) node {$a_{30}$}  (1.5,-1.5) node {$\ddots$};
\draw (0.5,-2.5) node {$\vdots$};
\draw (0.5,-3.5) node {$a_{r+2,0}$};

\draw (-0.25,0.5) node {$j$} (-0.5,0.25) node {$i$};
\draw (-0.15,0.15) -- (-0.55,0.55);
\end{tikzpicture}
\caption{Coefficients of the potential appearing in the expansion of the propagator.}
\label{diag_rhs}
\end{figure}
\begin{equation}
\bigg\lbrace a_{20}, \dots , a_{2,r}, a_{30}, \dots , a_{3,r-1}, \dots , a_{4,r-2}, \dots, a_{r+2,0}   \bigg\rbrace \,,
\end{equation}
(cf. figure \ref{diag_rhs}). Most of these coefficients have however already been accounted for on the left hand sides c.f. (\ref{coeff_lhs}). The only ones not counted yet are 
\begin{figure}
\centering
\begin{tikzpicture}[>=stealth]
\foreach \i in {0,...,6} {
\draw[pattern=north east lines] (0+\i,0) rectangle (1+\i,-1); }

\foreach \i in {0,...,5} {
\draw[pattern=north east lines] (0+\i,-1) rectangle (1+\i,-2); }

\foreach \i in {0,...,4} {
\draw[pattern=crosshatch] (0+\i,-2) rectangle (1+\i,-3); }
\draw[pattern=north west lines] (5,-2) rectangle (6,-3);

\foreach \i in {0,...,3} {
\draw[pattern=crosshatch] (0+\i,-3) rectangle (1+\i,-4); }
\draw[pattern=north west lines] (4,-3) rectangle (5,-4);

\foreach \i in {0,...,2} {
\draw[pattern=crosshatch] (0+\i,-4) rectangle (1+\i,-5); }
\draw[pattern=north west lines] (3,-4) rectangle (4,-5);

\foreach \i in {0,...,1} {
\draw[pattern=crosshatch] (0+\i,-5) rectangle (1+\i,-6); }
\draw[pattern=north west lines] (2,-5) rectangle (3,-6);

\draw[pattern=crosshatch] (0,-6) rectangle (1,-7);
\draw[pattern=north west lines] (1,-6) rectangle (2,-7);
\draw[pattern=north west lines] (0,-7) rectangle (1,-8);

\draw (0.5,0.5) node {0}  (1.5,0.5) node {1}  (2.5,0.5) node {2} (3.5,0.5) node {$\cdots$}
(4.5,0.5) node {$r-1$}  (5.5,0.5) node {$r$}  (6.5,0.5) node {$r+1$};
\draw (-0.5,-0.5) node {0}  (-0.5,-1.5) node {1}  (-0.5,-2.5) node {2} (-0.5,-3.5) node {$\vdots$}
(-0.5,-4.5) node {$r-1$}  (-0.5,-5.5) node {$r$}  (-0.5,-6.5) node {$r+1$}  (-0.5,-7.5) node {$r+2$};

\draw (-0.25,0.5) node {$j$} (-0.5,0.25) node {$i$};
\draw (-0.15,0.15) -- (-0.55,0.55);
\end{tikzpicture}
\caption{All the coefficients of the potential appearing on both sides of the equations.}
\label{diag_combined}
\end{figure}
\begin{equation}
\bigg\lbrace a_{2,r}, a_{3,r-1}, a_{4,r-2}, \dots, a_{r+2,0}   \bigg\rbrace \,,
\end{equation}
(cf. figure \ref{diag_combined}) which precisely add up to a further $r+1$ coefficients. We also must include another two coefficients, namely $\eta$ and $d_f$. 
Finally, since $\gamma$ is in general some function of $\chi$ it is easy to see that
\begin{equation}
\frac{d^r}{d \bar\chi^r} \, \bar{\gamma} \propto \frac{d^r}{d \bar\chi^r} \, \left( \frac{f'}{f}\right) \subseteq
\bigg\lbrace f, f', \dots, f^{(r+1)}   \bigg\rbrace \,,
\end{equation}
which gives us an additional $(r+2)$ coefficients from the Taylor expansion of $f$. The total number of coefficients is then given by
\begin{equation}
\label{number_coeff_general}
n_{\text{coeff}}(r) = n_{\text{lhs}} + (r+1) + (r+2) + 2  = \frac{1}{2} \, r^2 + \frac{9}{2} \, r + 8 \,.
\end{equation}

From \eqref{number_eqns} we see that for large $r$ the number of equations $\sim r^2$, while from \eqref{number_coeff_general} the number of coefficients only goes for large $r$ as $\sim r^2/2$. There are therefore asymptotically twice as many equations as coefficients, as already discussed in the Introduction. 
Equating the number of equations and coefficients yields the positive solution
\begin{equation}
r = 5.3 \,.
\end{equation}
Therefore the number of equations exceeds the number of coefficients for the first time at order $r=6$. If there is to be a conflict between the existence of ($k$-)fixed points and background independence generically we would expect this to become evident at about this level.  Equally, if there is no conflict between background independence and the existence of ($k$-)fixed points then from this level onwards some equations become redundant (\ie they provide constraints that are automatically satisfied once the other equations are obeyed). In the limit $r\to\infty$ fully half of the equations must become redundant if ($k$-)fixed points are to be consistent with background independence.

\section{Summary, discussion and conclusions}\label{sec:conclusions}

If we construct the non-perturbative flow equation for quantum gravity by introducing a cutoff defined through a background metric then independence from this artificial metric can only be achieved if the appropriate modified split Ward identity is obeyed. However even if it is obeyed, background independence is guaranteed only in the limit $k\to0$. RG properties on the other hand are defined at intermediate scales $k$. There is therefore the potential for conflict in this formulation between RG notions such as fixed points, and the requirement of background independence. Examples of such conflicts were uncovered in ref. \cite{Dietz:2015owa}.

\begin{table}[ht]
\begin{center}
\begin{tabular}{|c||c|c|c||c|c|c|c|c|c|c|}
\hline
 & \multicolumn{3}{c||}{parametrisation $f$} & \multicolumn{2}{c|}{cutoff profile $r$} \\
 \hline\hline
$\eta$ & type & $d_f$ & runs & power-law & not power-law \\
\hline \hline
\multirow{3}{*}{$\ne0$} & not power-law & any & yes & \xcancel{FP}\phantom{$=$} $\widehat{\rm FP}$ & \cellcolor{red!30} \\
\cline{2-5}
 & 
   power law & $\ne\rho\eta/2$ & yes & FP $\ne$ $\widehat{\rm FP}$ &  \cellcolor{red!30} incompatible \\
\cline{3-5}
 &     $f=\chi^\rho$    & $=\rho\eta/2$ & no & FP $=$ $\widehat{\rm FP}$ & \cellcolor{red!30} \\
\hline  
\multirow{2}{*}{$=0$} & \multirow{2}{*}{any} & $\ne0$ & yes & \multicolumn{2}{c|}{\multirow{2}{*}{FP $=$ $\widehat{\rm FP}$}} 
\\ 
\cline{3-4}
                   &         & $=0$    & no & \multicolumn{2}{c|}{}  \\
\hline
 
\end{tabular}
\end{center}
\caption{RG properties of the derivative expansion for conformally truncated gravity, when the msWI is also satisfied. The results depend on whether the conformal factor develops an anomalous dimension $\eta$, on the choice of cutoff profile $r$, and on how the metric is parametrised via $f$. Depending also on its dimension $d_f$, $f$ can contain a massive parameter, and thus run with $k$ when written in dimensionless terms, as listed in the table.
$\widehat{\rm FP}$ indicates that a background-independent description exists, while (\xcancel{FP}) FP  indicates that $k$-fixed points are (not) allowed; the (in)equality shows how these relate to the $\hk$-fixed points.}

\label{table:summary}
\end{table}

In this paper we have further investigated these issues. Our findings, together with those of ref. \cite{Dietz:2015owa}, are summarised in table \ref{table:summary}.\footnote{For power law cutoff  $r(z)=z^{-n}$, $d_f = 2-\eta/(n+2)$ is excluded \cite{Dietz:2015owa}, and from sec. \ref{sec:required-cutoff} when $\eta=0$,  $d_f=2$ is excluded for any cutoff profile: in these cases the cutoff term is independent of $k$.} The first question that needs to be addressed is whether the msWI, $\mathcal{W}=0$, is compatible with the exact RG flow equation, \ie such that $\dot{\mathcal{W}}=0$ then follows. At the exact level, this latter compatibility is guaranteed since they are both identities derived from the partition function (see also sec. \ref{sec:exact}). Within the derivative expansion approximation of conformally truncated gravity considered in ref. \cite{Dietz:2015owa} (reviewed in sec. \ref{sec:review}), we have shown  in secs. \ref{sec:compatibility-at-d2} and \ref{sec:required-cutoff}, that this compatibility follows if and only if either $\eta=0$ or cutoff profile is power law. In sec. \ref{sec:exact-vs-derivatives}, we saw precisely why the derivative expansion breaks compatibility in general and why these special cases restore it. We argued in sec. \ref{sec:incompatibility} that if the equations are incompatible they are overconstrained since there are then an infinite number of secondary constraints, and thus not even $t$-dependent solutions can exist. We confirmed this latter conclusion by example in sec. \ref{sec:incompatible-no-solns}. In sec. \ref{sec:truncations}, we also saw that the fixed point equations and Ward identities together generically overconstrain the system when expanded in terms of vertices beyond the six-point level.


Even if the equations are compatible the msWI can still forbid fixed points. In sec. \ref{sec:forbids} the reason was laid out particularly clearly. The Ward identity
\be 
\frac{\partial \bar V}{\partial \bar\chi} - \frac{\partial \bar V}{\partial \bar\varphi} + \bar \gamma \, \bar V = \bar \gamma
\int_0^{\infty} d\hat p \, \hat p^{d-1} \, \frac{r - \frac{1}{d} \, \hat p \, r'}{\hat p^2 + r - \partial^2_{\bar\varphi}\bar V} 
\ee
(which is compatible for power-law $r$),
forces the effective potential $\bV$ to depend on $t$ through
\be 
\bar{\gamma} = \frac{d}{2}\frac{\partial}{\partial\bc}\ln\bar{f}\left({\rm e}^{\eta t/2}\mu^{\eta/2}\bc\right)\,,
\ee
whenever this dimensionless combination is similarly forced to be $t$-dependent. For example  we see that fixed points with respect to $k$ are forbidden for exponential parametrisations $f(\phi)= \exp(\phi)$  if the field grows a non-zero anomalous dimension.  It is clear that the reasons for this conflict are general and not tied to the derivative expansion of the conformally truncated model \textit{per se}. Therefore this issue could provide important constraints for example on the exponential parametrisations recently advocated in the literature \cite{Demmel:2015zfa,Eichhorn:2013xr,Eichhorn:2015bna,Nink:2014yya,Percacci:2015wwa,Labus:2015ska,Ohta:2015efa,Gies:2015tca,Dona:2015tnf}.

For full quantum gravity, such conflict between $k$-fixed points and background independence may also show up clearly in a vertex expansion, as discussed in \ref{sec:truncations}, or generically it may not become visible until the six-point level. However for full quantum gravity, if we are to follow the standard procedure, we must also fix the gauge. The original msWI, which formally expresses background independence before gauge fixing, will no longer be compatible with the flow equation. Instead we must use the appropriate version which has  contributions from the background dependence of the gauge fixing and ghost terms as well as from cutoff terms for the ghost action itself. However background independence is then only restored in the limit $k\to0$ after going ``on-shell'' (assuming such an appropriate property can be defined). This last step is required to recover quantities that are independent of the gauge fixing. If we are to continue with a flow equation for a Legendre effective action \cite{Wetterich:1992,Morris:1993} then to get around this obstruction, the Vilkovisky-DeWitt covariant effective action seems called for \cite{Branchina:2003ek,Donkin:2012ud,Demmel:2014hla,Safari:2015dva}, with the msWI replaced by the corresponding modified Nielsen identities where the r\^ole of the background field is played by the ``base point'' \cite{Pawlowski:2003sk}.

Returning to the present study, it seems surely significant that  
whenever the msWI equations  are actually compatible with the flow equations, it is possible to combine them and thus uncover background independent variables, including a background independent notion of scale, $\hk$. These are not only independent of $\chi$ but also independent of the parametrisation $f$. Of course such an underlying description has only been shown in this $\mathcal{O}(\partial^2)$ approximation and in conformally truncated gravity, and one might doubt that this happy circumstance could be generalised to full quantum gravity, and not only for the reasons outlined above. However we also saw in sec. \ref{sec:counting} that if modified Ward identities are to be compatible with the flow equations then in terms of vertices, the information they contain becomes highly redundant at sufficiently high order (the six-point level in our case). This in itself suggests the existence of a simpler description. Finally very recently a formulation for non-perturbative RG has been proposed where computations can be made without ever introducing a background metric (or gauge fixing) \cite{Morris:2016nda}.

\section*{Acknowledgments}
TRM acknowledges support from STFC through Consolidated Grant ST/L000296/1. ZS acknowledges support through an STFC studentship. 
PL acknowledges support through Erasmus+ EU Grant and hospitality of Southampton University while this work was being completed.

\bibliographystyle{hunsrt}
\bibliography{references}

\begin{thebibliography}{10}

\bibitem{Reuter:1996}
M.~Reuter.
\newblock {Nonperturbative evolution equation for quantum gravity}.
\newblock {\em Phys.Rev.}, D57:971--985, 1998, hep-th/9605030.

\bibitem{Reuter:2012}
Martin Reuter and Frank Saueressig.
\newblock {Quantum Einstein Gravity}.
\newblock {\em New J.Phys.}, 14:055022, 2012, 1202.2274.

\bibitem{Percacci:2011fr}
Roberto Percacci.
\newblock {A Short introduction to asymptotic safety}.
\newblock 2011, 1110.6389.

\bibitem{Niedermaier:2006wt}
Max Niedermaier and Martin Reuter.
\newblock {The Asymptotic Safety Scenario in Quantum Gravity}.
\newblock {\em Living Rev.Rel.}, 9:5--173, 2006.

\bibitem{Nagy:2012ef}
Sandor Nagy.
\newblock {Lectures on renormalization and asymptotic safety}.
\newblock 2012, 1211.4151.

\bibitem{Litim:2011cp}
Daniel~F. Litim.
\newblock {Renormalisation group and the Planck scale}.
\newblock {\em Phil.Trans.Roy.Soc.Lond.}, A369:2759--2778, 2011, 1102.4624.

\bibitem{Pawlowski:2005xe}
Jan~M. Pawlowski.
\newblock {Aspects of the functional renormalisation group}.
\newblock {\em Annals Phys.}, 322:2831--2915, 2007, hep-th/0512261.

\bibitem{Litim:2002hj}
Daniel~F. Litim and Jan~M. Pawlowski.
\newblock {Wilsonian flows and background fields}.
\newblock {\em Phys.Lett.}, B546:279--286, 2002, hep-th/0208216.

\bibitem{Bridle:2013sra}
I.~Hamzaan Bridle, Juergen~A. Dietz, and Tim~R. Morris.
\newblock {The local potential approximation in the background field
  formalism}.
\newblock {\em JHEP}, 1403:093, 2014, 1312.2846.

\bibitem{Reuter:1997gx}
M.~Reuter and C.~Wetterich.
\newblock {Gluon condensation in nonperturbative flow equations}.
\newblock {\em Phys.Rev.}, D56:7893--7916, 1997, hep-th/9708051.

\bibitem{Litim:1998nf}
Daniel~F. Litim and Jan~M. Pawlowski.
\newblock {On gauge invariant Wilsonian flows}.
\newblock pages 168--185, 1998, hep-th/9901063.

\bibitem{Litim:2002ce}
Daniel~F. Litim and Jan~M. Pawlowski.
\newblock {Renormalization group flows for gauge theories in axial gauges}.
\newblock {\em JHEP}, 0209:049, 2002, hep-th/0203005.

\bibitem{Manrique:2009uh}
Elisa Manrique and Martin Reuter.
\newblock {Bimetric Truncations for Quantum Einstein Gravity and Asymptotic
  Safety}.
\newblock {\em Annals Phys.}, 325:785--815, 2010, 0907.2617.

\bibitem{Manrique:2010mq}
Elisa Manrique, Martin Reuter, and Frank Saueressig.
\newblock {Matter Induced Bimetric Actions for Gravity}.
\newblock {\em Annals Phys.}, 326:440--462, 2011, 1003.5129.

\bibitem{Manrique:2010am}
Elisa Manrique, Martin Reuter, and Frank Saueressig.
\newblock {Bimetric Renormalization Group Flows in Quantum Einstein Gravity}.
\newblock {\em Annals Phys.}, 326:463--485, 2011, 1006.0099.

\bibitem{Dietz:2015owa}
Juergen~A. Dietz and Tim~R. Morris.
\newblock {Background independent exact renormalization group for conformally
  reduced gravity}.
\newblock {\em JHEP}, 1504:118, 2015, 1502.07396.

\bibitem{Safari:2015dva}
Mahmoud Safari.
\newblock {Splitting Ward identity}.
\newblock 2015, 1508.06244.

\bibitem{Morris:2016nda}
Tim~R. Morris and Anthony W.~H. Preston.
\newblock {Manifestly diffeomorphism invariant classical Exact Renormalization
  Group}.
\newblock 2016, 1602.08993.

\bibitem{Dietz2016}
Juergen~A. Dietz and Tim~R. Morris.
\newblock Background independent conformally reduced gravity and asymptoic
  safety.
\newblock {\em to appear}, 2016.

\bibitem{Morris:1994ie}
Tim~R. Morris.
\newblock {Derivative expansion of the exact renormalization group}.
\newblock {\em Phys.Lett.}, B329:241--248, 1994, hep-ph/9403340.

\bibitem{Morris:1994jc}
Tim~R. Morris.
\newblock {The Renormalization group and two-dimensional multicritical
  effective scalar field theory}.
\newblock {\em Phys.Lett.}, B345:139--148, 1995, hep-th/9410141.

\bibitem{Morris:1998}
Tim~R. Morris.
\newblock {Elements of the continuous renormalization group}.
\newblock {\em Prog.Theor.Phys.Suppl.}, 131:395--414, 1998, hep-th/9802039.

\bibitem{opt1}
Daniel~F. Litim.
\newblock {Optimization of the Exact Renormalization Group}.
\newblock {\em Phys.Lett.}, B486:92--99, {(2000)}, hep-th/0005245.

\bibitem{opt3}
Daniel~F. Litim.
\newblock {Mind the Gap}.
\newblock {\em Int.J.Mod.Phys.}, A16:2081--2088, {(2001)}, hep-th/0104221.

\bibitem{Morris:2005ck}
Tim~R. Morris.
\newblock {Equivalence of local potential approximations}.
\newblock {\em JHEP}, 0507:027, 2005, hep-th/0503161.

\bibitem{Morris1999}
Tim~R. Morris and John~F. Tighe.
\newblock {Convergence of derivative expansions of the renormalization group}.
\newblock {\em JHEP}, 08:007, 1999, hep-th/9906166.

\bibitem{Morris2001}
Tim~R. Morris and John~F. Tighe.
\newblock {Convergence of derivative expansions in scalar field theory}.
\newblock {\em Int. J. Mod. Phys.}, A16:2095--2100, 2001, hep-th/0102027.

\bibitem{Machado:2009ph}
Pedro~F. Machado and R.~Percacci.
\newblock {Conformally reduced quantum gravity revisited}.
\newblock {\em Phys.Rev.}, D80:024020, 2009, 0904.2510.

\bibitem{Bonanno:2012dg}
Alfio Bonanno and Filippo Guarnieri.
\newblock {Universality and Symmetry Breaking in Conformally Reduced Quantum
  Gravity}.
\newblock {\em Phys.Rev.}, D86:105027, 2012, 1206.6531.

\bibitem{Wetterich:1992}
Christof Wetterich.
\newblock {Exact evolution equation for the effective potential}.
\newblock {\em Phys.Lett.}, B301:90--94, 1993.

\bibitem{Morris:1993}
Tim~R. Morris.
\newblock {The Exact renormalization group and approximate solutions}.
\newblock {\em Int.J.Mod.Phys.}, A9:2411--2450, 1994, hep-ph/9308265.

\bibitem{Reuter:2008wj}
Martin Reuter and Holger Weyer.
\newblock {Background Independence and Asymptotic Safety in Conformally Reduced
  Gravity}.
\newblock {\em Phys.Rev.}, D79:105005, 2009, 0801.3287.

\bibitem{Reuter:2009kq}
Martin Reuter and Holger Weyer.
\newblock {The Role of Background Independence for Asymptotic Safety in Quantum
  Einstein Gravity}.
\newblock {\em Gen.Rel.Grav.}, 41:983--1011, 2009, 0903.2971.

\bibitem{Litim1998}
Daniel~F. Litim and Jan~M. Pawlowski.
\newblock {Flow equations for Yang-Mills theories in general axial gauges}.
\newblock {\em Phys. Lett.}, B435:181--188, 1998, hep-th/9802064.

\bibitem{Litim1999}
Daniel~F. Litim and Jan~M. Pawlowski.
\newblock {On gauge invariance and Ward identities for the Wilsonian
  renormalization group}.
\newblock {\em Nucl. Phys. Proc. Suppl.}, 74:325--328, 1999, hep-th/9809020.

\bibitem{Dirac-primary}
Paul A.~M. Dirac.
\newblock {\em Lectures on quantum mechanics}.
\newblock Belfer Graduate School of Science Monographs Series 2. Belfer
  Graduate School of Science, New York, reprinted by Dover, 2001.

\bibitem{Dirac1950}
P.~A.~M. Dirac.
\newblock Generalized hamiltonian dynamics.
\newblock {\em Canad. J. Math.}, 2(0):129--148, Jan 1950.

\bibitem{Demmel:2015zfa}
Maximilian Demmel and Andreas Nink.
\newblock {Connections and geodesics in the space of metrics}.
\newblock {\em Phys. Rev.}, D92(10):104013, 2015, 1506.03809.

\bibitem{Eichhorn:2013xr}
Astrid Eichhorn.
\newblock {On unimodular quantum gravity}.
\newblock {\em Class. Quant. Grav.}, 30:115016, 2013, 1301.0879.

\bibitem{Eichhorn:2015bna}
Astrid Eichhorn.
\newblock {The Renormalization Group flow of unimodular f(R) gravity}.
\newblock 2015, 1501.05848.

\bibitem{Nink:2014yya}
Andreas Nink.
\newblock {Field Parametrization Dependence in Asymptotically Safe Quantum
  Gravity}.
\newblock 2014, 1410.7816.

\bibitem{Percacci:2015wwa}
Roberto Percacci and Gian~Paolo Vacca.
\newblock {Search of scaling solutions in scalar-tensor gravity}.
\newblock 2015, 1501.00888.

\bibitem{Labus:2015ska}
Peter Labus, Roberto Percacci, and Gian~Paolo Vacca.
\newblock {Asymptotic safety in $O(N)$ scalar models coupled to gravity}.
\newblock {\em Phys. Lett.}, B753:274--281, 2016, 1505.05393.

\bibitem{Ohta:2015efa}
Nobuyoshi Ohta, Roberto Percacci, and Gian~Paolo Vacca.
\newblock {A flow equation for f(R) gravity and some of its exact solutions}.
\newblock 2015, 1507.00968.

\bibitem{Gies:2015tca}
Holger Gies, Benjamin Knorr, and Stefan Lippoldt.
\newblock {Generalized Parametrization Dependence in Quantum Gravity}.
\newblock {\em Phys. Rev.}, D92(8):084020, 2015, 1507.08859.

\bibitem{Dona:2015tnf}
Pietro Don{\`a}, Astrid Eichhorn, Peter Labus, and Roberto Percacci.
\newblock {Asymptotic safety in an interacting system of gravity and scalar
  matter}.
\newblock {\em Phys. Rev.}, D93(4):044049, 2016, 1512.01589.

\bibitem{Branchina:2003ek}
Vincenzo Branchina, Krzysztof~A. Meissner, and Gabriele Veneziano.
\newblock {The Price of an exact, gauge invariant RG flow equation}.
\newblock {\em Phys.Lett.}, B574:319--324, 2003, hep-th/0309234.

\bibitem{Donkin:2012ud}
Ivan Donkin and Jan~M. Pawlowski.
\newblock {The phase diagram of quantum gravity from diffeomorphism-invariant
  RG-flows}.
\newblock 2012, 1203.4207.

\bibitem{Demmel:2014hla}
Maximilian Demmel, Frank Saueressig, and Omar Zanusso.
\newblock {RG flows of Quantum Einstein Gravity in the linear-geometric
  approximation}.
\newblock 2014, 1412.7207.

\bibitem{Pawlowski:2003sk}
Jan~M. Pawlowski.
\newblock {Geometrical effective action and Wilsonian flows}.
\newblock 2003, hep-th/0310018.

\end{thebibliography}

\end{document}